\documentclass[10pt]{article}
\usepackage{fullpage}
\usepackage{times}
\usepackage{graphicx}
\usepackage{amssymb,amsmath,amstext}

\newcommand{\Te}{\text{eff}}
\newcommand{\Tv}{\text{V}}
\newcommand{\TwoD}{\text{2D}}
\newcommand{\Ceff}[1]{\ensuremath{C^{\Te}_{#1}}}
\newcommand{\Eeff}[1]{\ensuremath{E^{\Te}_{#1}}}
\newcommand{\Neff}[1]{\ensuremath{\nu^{\Te}_{#1}}}
\newcommand{\Geff}[1]{\ensuremath{G^{\Te}_{#1}}}
\newcommand{\Avg}[1]{\ensuremath{\left<#1\right>}}
\newcommand{\VAvg}[1]{\ensuremath{\Avg{#1}_{\Tv}}}
\newcommand{\Sigx}{\sigma_{11}}
\newcommand{\Sigy}{\sigma_{22}}
\newcommand{\Sigz}{\sigma_{33}}
\newcommand{\ASigx}{\VAvg{\Sigx}}
\newcommand{\ASigy}{\VAvg{\Sigy}}
\newcommand{\ASigz}{\VAvg{\Sigz}}
\newcommand{\Tauxy}{\sigma_{12}}
\newcommand{\Tauyz}{\sigma_{23}}
\newcommand{\Tauzx}{\sigma_{31}}
\newcommand{\ATauxy}{\VAvg{\Tauxy}}
\newcommand{\ATauyz}{\VAvg{\Tauyz}}
\newcommand{\ATauzx}{\VAvg{\Tauzx}}
\newcommand{\Epsx}{\epsilon_{11}}
\newcommand{\Epsy}{\epsilon_{22}}
\newcommand{\Epsz}{\epsilon_{33}}
\newcommand{\AEpsx}{\VAvg{\Epsx}}
\newcommand{\AEpsy}{\VAvg{\Epsy}}
\newcommand{\AEpsz}{\VAvg{\Epsz}}
\newcommand{\Gamxy}{\epsilon_{12}}
\newcommand{\Gamyz}{\epsilon_{23}}
\newcommand{\Gamzx}{\epsilon_{31}}
\newcommand{\AGamxy}{\VAvg{\Gamxy}}
\newcommand{\AGamyz}{\VAvg{\Gamyz}}
\newcommand{\AGamzx}{\VAvg{\Gamzx}}

\begin{document}

  \title{Effective elastic moduli of polymer bonded explosives from 
  finite element simulations}

  \author{Biswajit Banerjee\footnote{E-mail: 
  banerjee@eng.utah.edu~~ Fax: (801)-585-00396}\\
  Dept. of Mechanical Engineering, University of Utah, \\
  50 S Central Campus Drive, Salt Lake City, UT 84112, USA.}

  \maketitle

  \raggedright
  \section*{Abstract}
  Finite element analysis has been used successfully to estimate the effective
  properties of many types of composites.  The prediction of effective elastic
  moduli of polymer-bonded explosives provides a new challenge.  These 
  particulate composites contain extremely high volume fractions of explosive 
  particles ($>$ 0.90).  At room temperature and higher, the Young's modulus 
  of the particles can be 20,000 times that of the binder.  Under these 
  conditions, rigorous bounds and analytical approximations for effective 
  elastic properties predict values that are orders of magnitude different 
  from the experimental values.  In this work, an approach is presented that 
  can be used to predict three-dimensional effective elastic moduli from 
  two-dimensional finite element simulations.  The approach is validated by 
  comparison with differential effective medium estimates and 
  three-dimensional finite element simulations.  The two-dimensional finite 
  element approach has been used to determine the properties of models of 
  polymer bonded explosives and PBX 9501 in particular, containing high volume 
  fractions of circular and square particles with high modulus 
  contrasts.  Results show that estimates of effective elastic properties from 
  two-dimensional finite element calculations are close to the values predicted
  by the differential effective medium approach for a large range of volume 
  fractions and modulus contrasts.  Two- and three-dimensional finite element 
  estimates for volume fractions from 0.70 to 0.90 and found not to differ 
  considerably.  Simulations of models of polymer bonded explosives and PBX 
  9501 show that the microstructure, the amount of discretization, and the 
  type of element used play a considerable role in determining the value 
  predicted by finite element simulations.  The effective elastic moduli of 
  PBX 9501 predicted by finite element calculations can vary from 200 MPa to 
  10,000 MPa depending on the microstructure and level of discretization 
  used.  The results also suggest that if a microstructure can be 
  found that accurately predicts the elastic properties of PBX 9501 at a 
  certain temperature and strain rate, then the same microstructure can be 
  used to predict elastic properties at other temperatures and strain rates.

  Keywords : particle-reinforced composites, microstructure, modeling, 
  elastic properties

  \section{Introduction}
  Experimental determination of the mechanical properties of polymer-bonded 
  explosives (PBXs) is hazardous and highly expensive.  An alternative to 
  experimentation is the determination of material properties using 
  micromechanics methods.  However, rigorous bounds and analytical approaches 
  predict values of elastic modulus that are orders of magnitude different 
  from the experimental values.  Hence, detailed numerical simulations are 
  required for the prediction of effective properties of PBXs.  The goal of 
  this work is to investigate if the elastic properties of PBXs can be 
  determined using finite element analyses of PBX microstructures.

  PBXs are particulate composites containing explosive particles and a 
  rubbery binder.  The particle size distributions in these materials and 
  the mechanical properties of the constituents are 
  required prior to finite element simulations of the microstructure of 
  these materials.  The size distribution of the explosive particles in
  PBXs can be obtained experimentally.  The mechanical properties of the 
  particles can be determined from molecular dynamics simulations.  Mechanical 
  testing of the rubbery binder does not involve any risk of 
  explosion.  Hence, mechanical properties of the binder can be determined 
  from laboratory tests.  

  The major challenges involved in finite element modeling of PBXs are the 
  high volume fraction of particles in PBXs ($f_p >$ 0.90) and the high 
  modulus contrast between particles and binder ($E_p/E_b = $ 10,000 to 
  20,000) at and above room temperature and at low strain rates.  Because 
  of the high volume fractions, microstructures are difficult to generate 
  and even more difficult to digitize from micrographs.  In addition, the 
  high modulus contrast leads to large stress concentrations at the 
  interface of particles and the binder.  These stress concentrations 
  cannot be simulated adequately without a high degree of mesh 
  discretization.  Hence, three-dimensional finite element models of these 
  materials require considerable computational power.  

  In this work, a procedure is outlined whereby two-dimensional finite 
  element simulations can be used to estimate the three-dimensional elastic 
  moduli.  In order to verify the accuracy of the approach, finite element 
  estimates on unit cells containing circular particles are compared with 
  differential effective medium estimates~\cite{Berry96,Markov00} of 
  effective elastic moduli for a wide range of volume fractions and modulus 
  contrasts.  Two-dimensional finite element estimates for selected 
  microstructures are then compared to three-dimensional estimates for 
  particle volume fractions of 0.70, 0.75 and 0.80.  

  Next, an appropriate model of PBX 9501 is determined from simulations of 
  manually generated and randomly generated microstructures based on the 
  particle size distribution of PBX 9501.  Finite element calculations are 
  then performed on this model to estimate the elastic moduli of PBX 9501 
  for various temperatures and strain rates.  Finally, the estimated 
  moduli are compared with experimental data to gauge the applicability
  of the approach to polymer bonded explosives.

  \section{Finite element approach}\label{sec:feApproach}
  Numerical determination of the effective properties of particulate 
  composites involves the calculation of the stress and strain fields for 
  a representative volume element (RVE) that simulates the microstructure 
  of the composite.  These stresses and strains are averaged over the 
  volume (V) of the RVE.  The effective elastic stiffness tensor \Ceff{ijkl} 
  of the composite is calculated from the tensor relation 
  \begin{equation}
     \int_V \sigma_{ij} dV = \Ceff{ijkl} \int_V \epsilon_{kl} dV, 
     \label{eq:sigepsFE}
  \end{equation}
  where $\sigma_{ij}$ are the stresses and $\epsilon_{ij}$ are the 
  strains.  

  The stress-strain relation for the RVE can be written in Voigt notation as
  \begin{equation}
    \begin{bmatrix} \ASigx \\ \ASigy \\ \ASigz \\ 
                    \ATauyz \\ \ATauzx \\ \ATauxy \end{bmatrix}  =
    \begin{bmatrix} \Ceff{11} & \Ceff{12} & \Ceff{13} & 0 & 0 & 0\\
                    \Ceff{12} & \Ceff{22} & \Ceff{23} & 0 & 0 & 0\\
                    \Ceff{13} & \Ceff{23} & \Ceff{33} & 0 & 0 & 0\\
                    0 & 0 & 0 & \Ceff{44} & 0 & 0 \\
                    0 & 0 & 0 & 0 & \Ceff{55} & 0 \\
                    0 & 0 & 0 & 0 & 0 & \Ceff{66} \end{bmatrix}
    \begin{bmatrix} \AEpsx \\ \AEpsy \\ \AEpsz \\ 
                    \AGamyz \\ \AGamzx \\ \AGamxy \end{bmatrix} 
    \label{eq:sigepsFEM1}
  \end{equation}
  where $\Avg{a}_V$ is the volume average of the quantity 
  $a$.  Equation~(\ref{eq:sigepsFEM1}) can also be inverted and written 
  in terms of the directional Young's moduli, Poisson's ratios and 
  shear moduli as
  \begin{equation}
    \begin{bmatrix} \AEpsx \\ \AEpsy \\ \AEpsz \\ 
                    2\AGamyz \\ 2\AGamzx \\ 2\AGamxy \end{bmatrix} =
    \begin{bmatrix} 1/\Eeff{11} & -\Neff{21}/\Eeff{11} & 
                                -\Neff{31}/\Eeff{11} & 0 & 0 & 0  \\ 
                    -\Neff{12}/\Eeff{22} & 1/\Eeff{22} & 
                                -\Neff{32}/\Eeff{22} & 0 & 0 & 0  \\
                    -\Neff{13}/\Eeff{33} & -\Neff{23}/\Eeff{33} & 1/\Eeff{33} &
                                 0 & 0 & 0  \\
                    0 & 0 & 0 & 1/\Geff{23} & 0 & 0  \\
                    0 & 0 & 0 & 0 & 1/\Geff{31} & 0  \\
                    0 & 0 & 0 & 0 & 0 & 1/\Geff{12} \end{bmatrix}
    \begin{bmatrix} \ASigx \\ \ASigy \\ \ASigz \\ 
                    \ATauyz \\ \ATauzx \\ \ATauxy \end{bmatrix} 
    \label{eq:sigepsFE2}
  \end{equation}
  where the directional Young's moduli \Eeff{11}, \Eeff{22}, and 
  \Eeff{33} differ by a small amount from the isotropic Young's 
  modulus $E_{\Te}$.  Similarly, the mean of the Poisson's ratios 
  \Neff{12}, \Neff{21}, \Neff{13}, \Neff{31}, \Neff{23}, and \Neff{32} is 
  assumed to be close to the isotropic Poisson's ratio $\nu_{\Te}$ and the 
  mean of the shear moduli \Geff{12}, \Geff{23}, and \Geff{31} is assumed to 
  be close to the isotropic shear modulus $G_{\Te}$.  
  The bulk mechanical response of a random particulate composite is 
  isotropic.  However, the size of a RVE of such a composite that can be 
  simulated numerically is necessarily much smaller than the bulk and may 
  not be isotropic.  In this work, it is assumed that the deviation 
  from isotropy of an RVE is small.  

  The determination of effective isotropic elastic modulus of particulate 
  composites using three-dimensional finite element analysis therefore 
  requires the determination of volume averaged stresses and strains under 
  appropriate boundary conditions.  Once these stresses and strains are known 
  for an RVE, equation~(\ref{eq:sigepsFE2}) can be used to determine the 
  effective elastic moduli of the RVE.  An investigation into effects of 
  boundary conditions revealed a uniaxial state of stress develops in a RVE 
  when a uniform boundary displacement is applied in one direction and 
  periodicity of boundary displacements is maintained.  This means that the 
  effective Young's modulus and Poisson's ratio in a particular direction 
  can be directly computed from equation~(\ref{eq:sigepsFE2}) by applying a 
  uniform displacement in that direction.  The effective isotropic Young's 
  modulus and Poisson's ratio can then be calculated by averaging the values 
  obtained in the three orthogonal directions from three simulations.  The 
  effective shear modulus can be calculated directly from the 
  effective isotropic Young's modulus and Poisson's ratio.

  The estimation of effective elastic properties from three-dimensional 
  simulations is straightforward in principle.  However, not only is it 
  difficult to generate RVEs containing high volume fractions of random 
  spherical particles, meshing such three-dimensional RVEs is 
  nontrivial.  Even if RVEs can be generated and meshed, solution of the 
  finite element system of equations can be prohibitive in terms of 
  computational cost.  A two-dimensional approach is described below that 
  requires relatively little computational expense.  This approach can 
  be used to arrive at reasonable estimates of three-dimensional effective 
  properties of particulate composites.

  In the two-dimensional simulations of particulate composites performed 
  in this study, a cross-section of the composite is chosen that contains 
  the same area fraction of particles as the volume fraction of particles 
  in the composite.  Finite element analyses are then performed on this 
  square RVE assuming a state of plane strain.  The stress-strain 
  equation~(\ref{eq:sigepsFE2}) can be written in planar or two-dimensional 
  form as 
  \begin{equation}
    \begin{bmatrix} \AEpsx \\ \AEpsy \\ 2\AGamxy \end{bmatrix} =
    \begin{bmatrix} 1/\widehat{\Eeff{11}} & 
                    -\widehat{\Neff{21}}/\widehat{\Eeff{11}} & 0  \\ 
                    -\widehat{\Neff{12}}/\widehat{\Eeff{22}} & 
                    1/\widehat{\Eeff{22}} & 0  \\
                    0 & 0 & 1/\Geff{12} \end{bmatrix}
    \begin{bmatrix} \ASigx \\ \ASigy \\ \ATauxy \end{bmatrix} 
    \label{eq:sigepsFE3}
  \end{equation}
  where $\widehat{\Eeff{11}}$, $\widehat{\Eeff{22}}$, $\widehat{\Neff{21}}$, 
  and $\widehat{\Neff{12}}$ are two-dimensional Young's moduli and Poisson's 
  ratios and different from their three-dimensional counterparts.  A brief 
  explanation of two-dimensional elastic moduli is given in 
  Appendix~\ref{app:2D3D}.  

  Uniform displacements are applied to the `1' and `2' directions of the 
  RVE and periodic displacements are enforced along the remainder of the 
  boundary to arrive at uniaxial average stresses and the corresponding 
  average strains in the RVE.  The effective moduli are calculated using 
  equation~(\ref{eq:sigepsFE3}) and the directional moduli are averaged to 
  arrive at a mean effective Young's modulus ($E^{2D}_{\Te}$) and a mean 
  effective Poisson's ratio ($\nu^{2D}_{\Te}$) for the RVE.  These moduli 
  can also be thought of as effective transverse moduli of a composite 
  containing unidirectional cylinders.

  It is assumed that the transverse moduli are equal to the two-dimensional 
  apparent moduli of an isotropic, homogeneous material subjected to plane 
  strain.  Since equation (\ref{eq:sigepsFE3}) is a two-dimensional 
  stress-strain law, the two-dimensional moduli have to be converted to 
  three-dimensional moduli using the relations~\cite{JunJas93}
  \begin{eqnarray}
     \nu_{\Te}  & = & \nu_{\Te}^{\TwoD}/(1+\nu_{\Te}^{\TwoD})\nonumber\\
     E_{\Te} & =  & E^{\TwoD}_{\Te} [1-(\nu_{\Te})^2],\label{eq:2Dto3D}
  \end{eqnarray}
  where $\nu_{\Te}$ is the effective Poisson's ratio in three dimensions and 
  $E_{\Te}$ is the three-dimensional effective Young's modulus.
  
  \section{Validation of approach}

  \subsection{Two-dimensional FEM vs. differential effective medium estimates}
\begin{figure}
    \begin{center}
     \scalebox{0.75}{\includegraphics{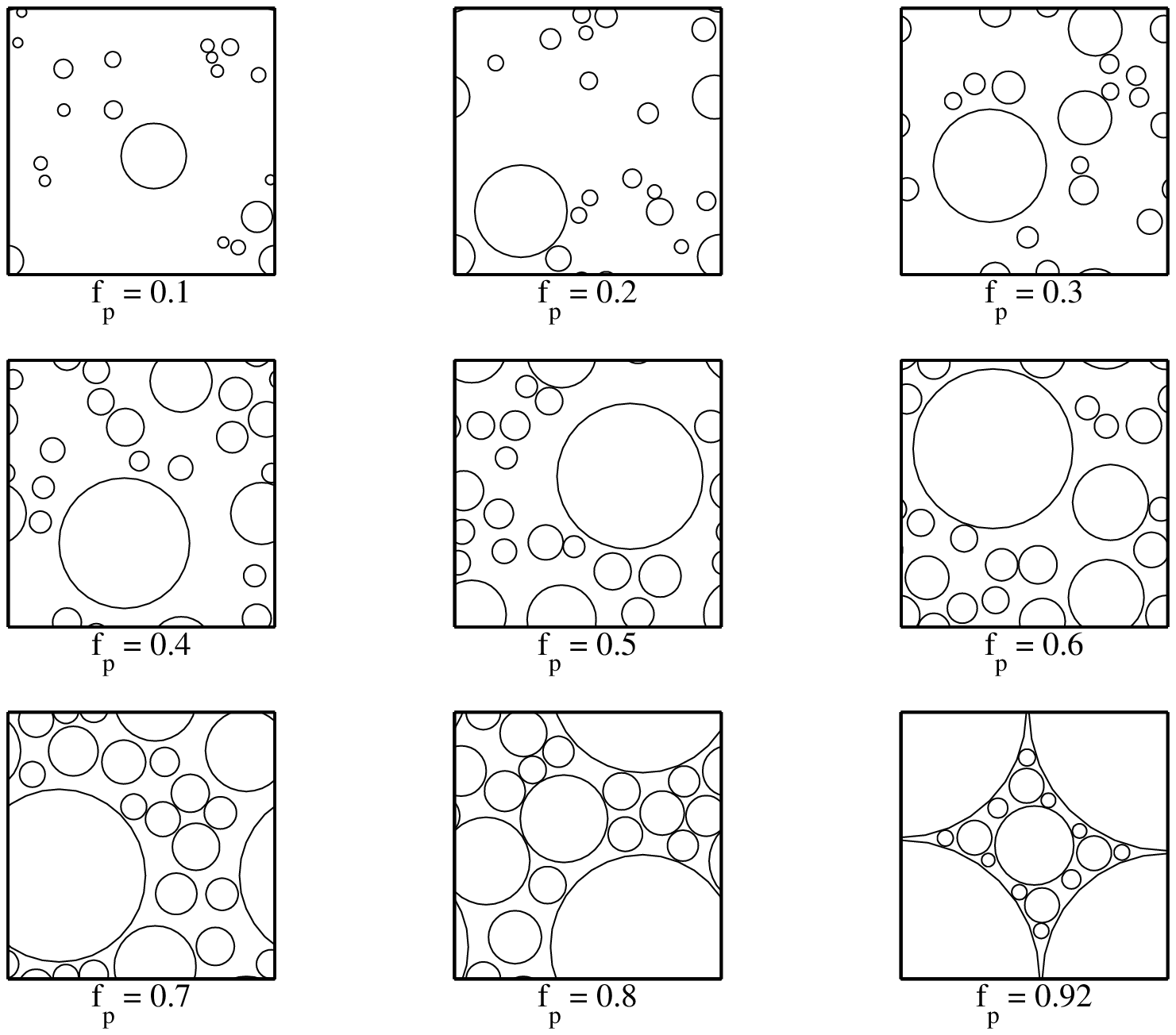}}
    \end{center}
     \caption{RVEs containing 10\% to 92\% by volume of circular particles.
              $f_p$ is the volume fraction of particles in a RVE.}
     \label{fig:nineRVEs}
  \end{figure}

  The differential effective medium (DEM) approach~\cite{Berry96,Markov00} 
  has been found to provide excellent estimates of elastic properties of 
  particulate composites~\cite{Markov00}.  To determine if the two-dimensional 
  finite element approach provides reasonable estimates of three-dimensional 
  effective elastic properties, finite element analyses were 
  performed on the two-dimensional RVEs containing 10\% to 92\% by volume 
  of circular particles that are shown in Figure~\ref{fig:nineRVEs}.  
  The particles were assigned a Young's modulus of 100,000 MPa and a 
  Poisson's ratio of 0.2.  The Young's modulus of the binder was varied 
  from 1 MPa to 10,000 MPa in multiples of 10 and the binder Poisson's ratio 
  was chosen to be 0.49. The effective properties predicted by the finite 
  element approach were then compared with DEM estimates.

\begin{figure}
    \begin{center}
     \scalebox{0.45}{\includegraphics{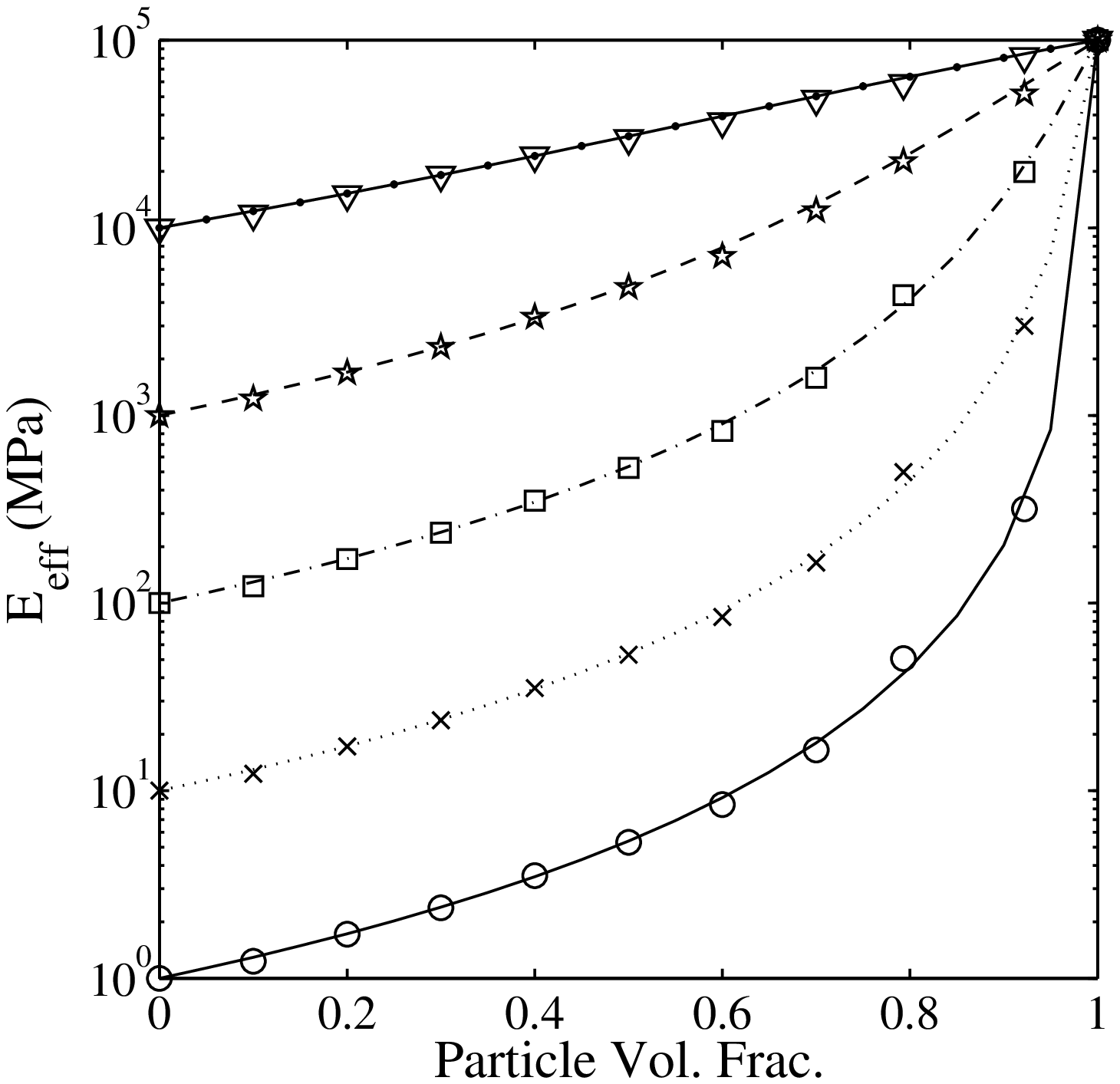}
                     \includegraphics{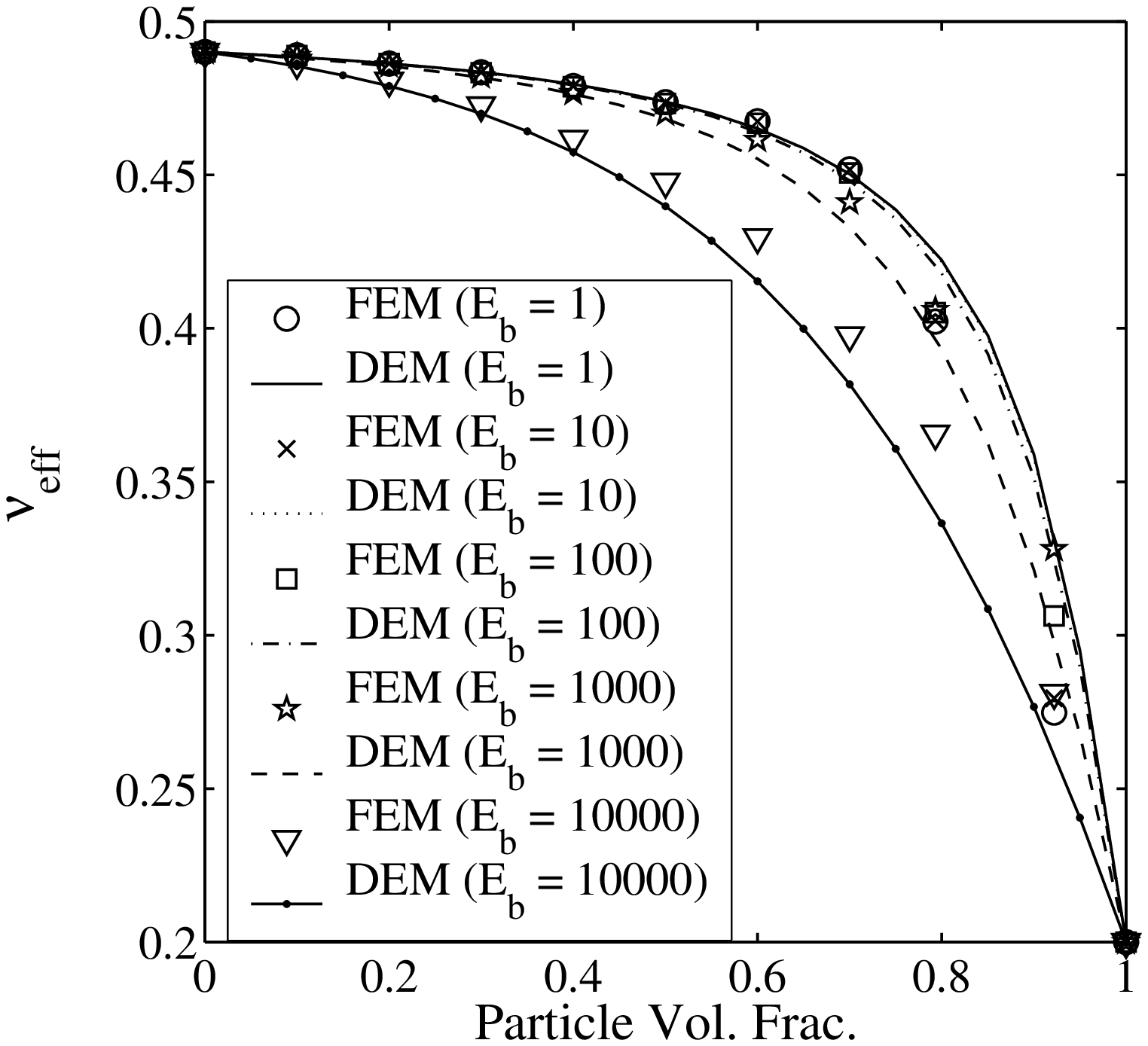}}\\
     (a) Young's Modulus.\hspace{1.4in}(b)Poisson's Ratio.
    \end{center}
     \caption{Comparison of finite element (FEM) and differential
              effective medium (DEM) predictions.  $E_b$ is the Young's 
              modulus of the binder.}
     \label{fig:DEMFEM}
  \end{figure}

  In the RVEs shown in Figure~\ref{fig:nineRVEs}, circular particles were 
  placed sequentially at random locations in each RVE such that no two 
  particles were in contact.  The particle size distribution in each RVE was 
  based on that of the dry blend of PBX 9501~\cite{Skid97}.  The planar moduli 
  obtained from the finite element calculations were converted into 
  three-dimensional moduli using equations~\ref{eq:2Dto3D}.  
  Figure~\ref{fig:DEMFEM}(a) shows effective three-dimensional Young's moduli 
  of the RVEs from finite element (FEM) and differential effective 
  medium (DEM) calculations.  Figure~\ref{fig:DEMFEM}(b) shows the 
  corresponding three-dimensional Poisson's ratios.  The solid lines in 
  the figures are lines of constant modulus contrast between particles and 
  binder and represent the effective elastic property calculated using 
  DEM.  The finite element results are shown as squares.  
 
  The FEM and the DEM predictions are seen to agree remarkably well for all 
  modulus contrasts and for all the simulated volume fractions of 
  particles, implying that the two-dimensional FEM approach described in 
  the previous section gives reasonable estimates of effective elastic 
  properties over a considerable range of volume fractions and modulus 
  contrasts.  

  \subsection{Two-dimensional vs. three-dimensional FEM estimates}
\begin{figure}
    \begin{center}
     $f_p = 0.7$\hspace{1.2in}$f_p = 0.75$\hspace{1.2in}$f_p = 0.8$\\
     \vspace{12pt}
     \scalebox{0.35}{\includegraphics{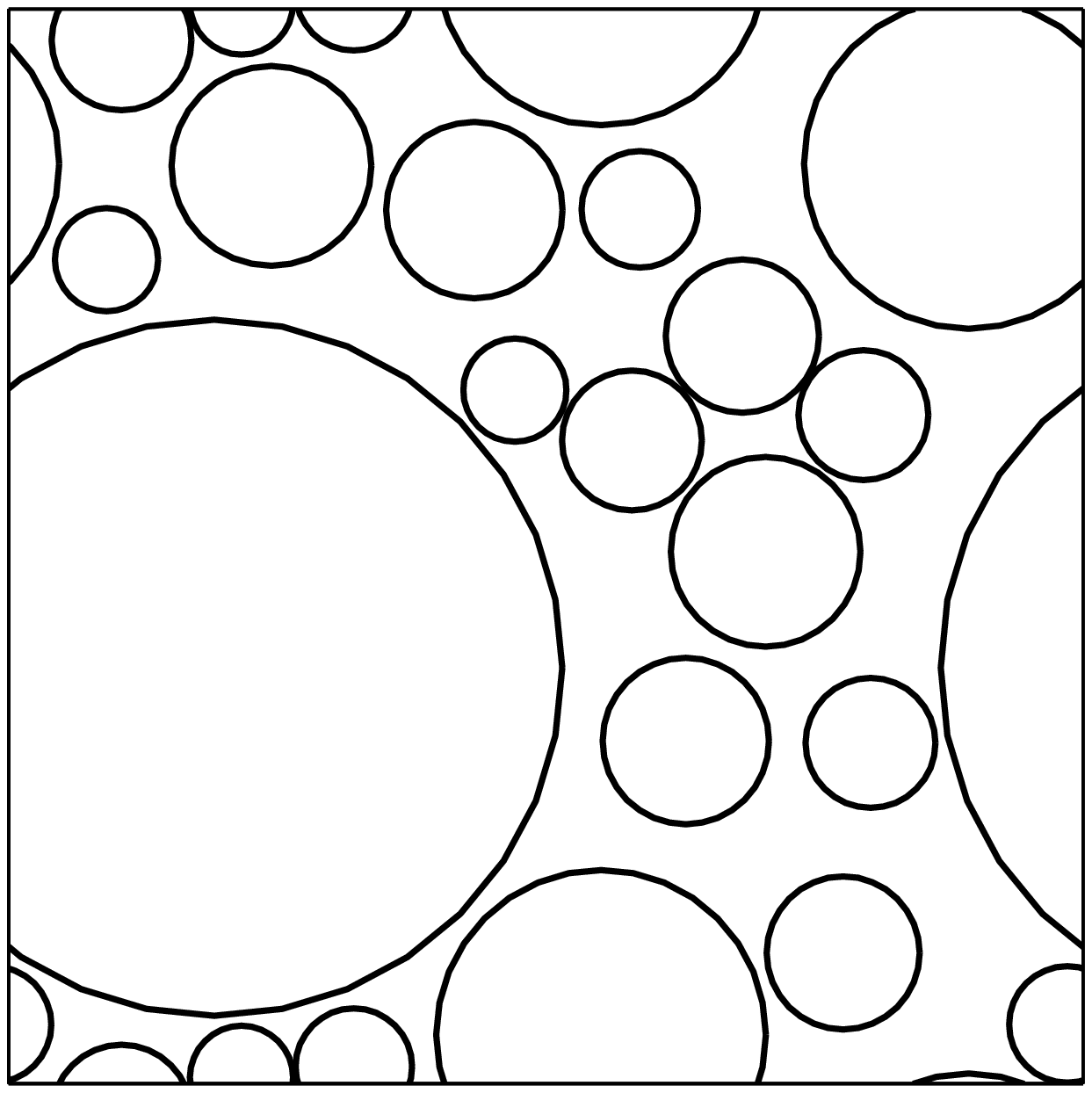}\hspace{24pt}
                     \includegraphics{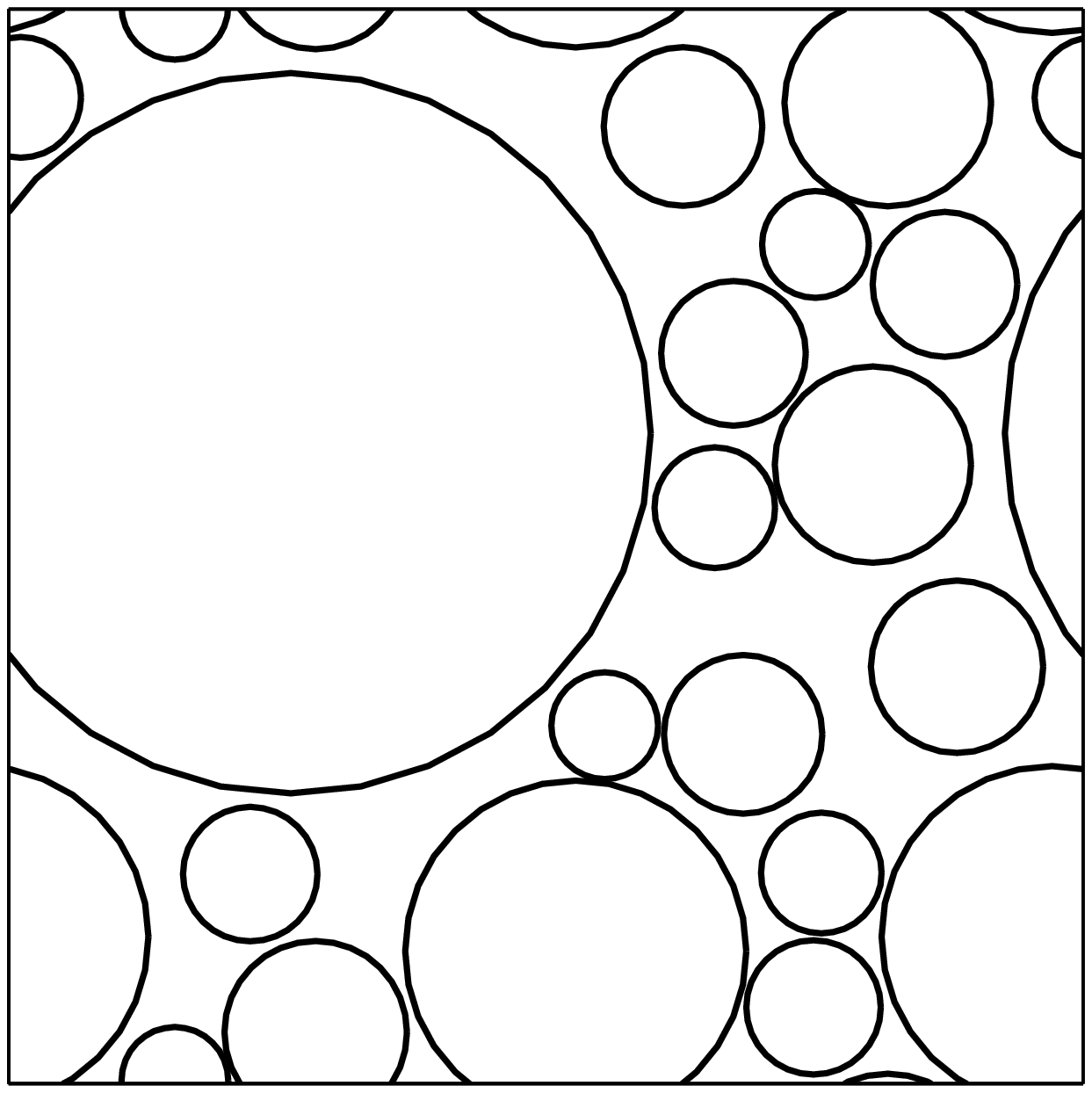}\hspace{24pt}
                     \includegraphics{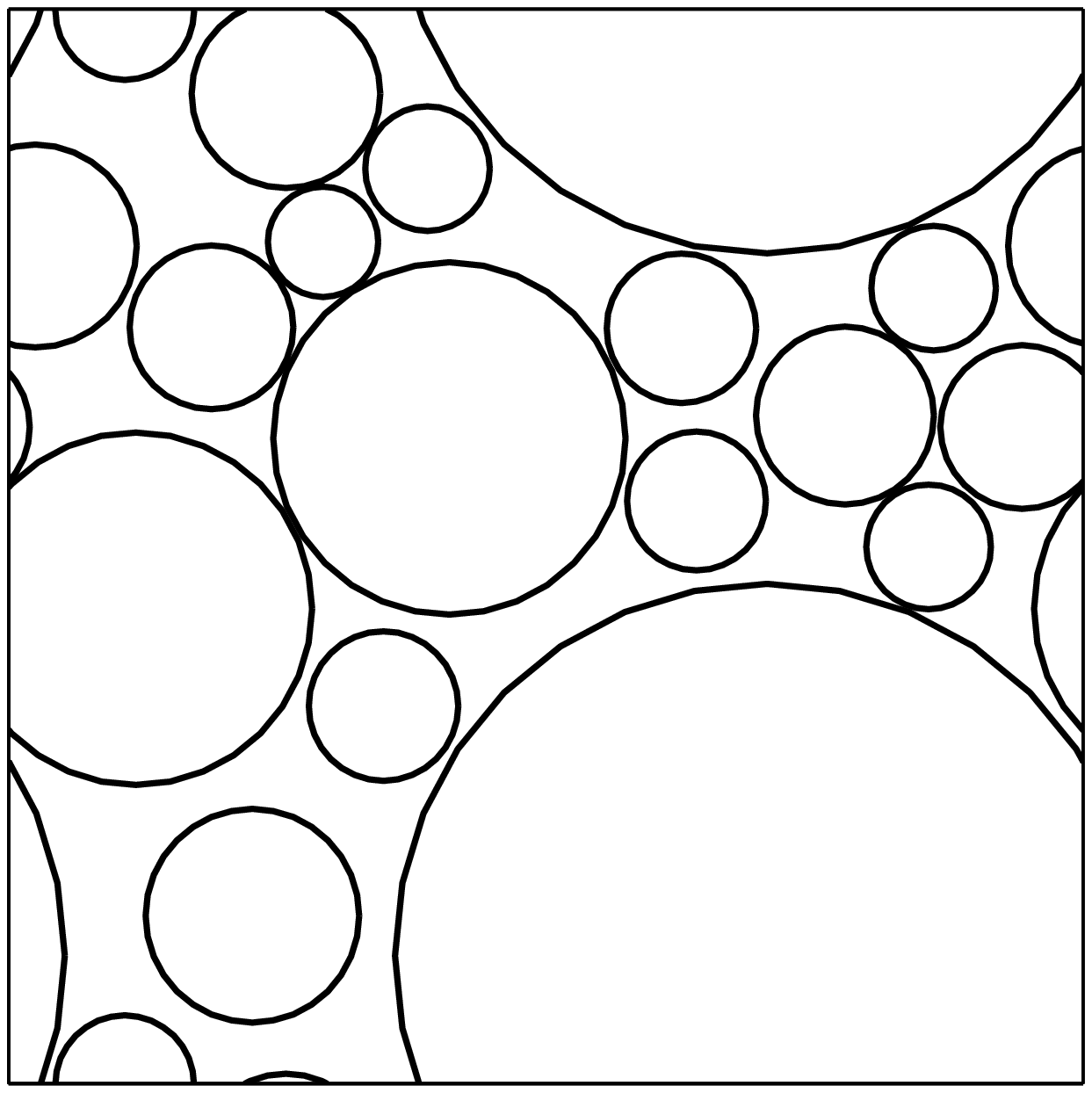}}\\
     (a) Two-dimensional RVEs.\\
     \vspace{24pt}
     $f_p = 0.7$\hspace{1.3in}$f_p = 0.75$\hspace{1.3in}$f_p = 0.8$\\
     \vspace{12pt}
     \scalebox{0.34}{\includegraphics{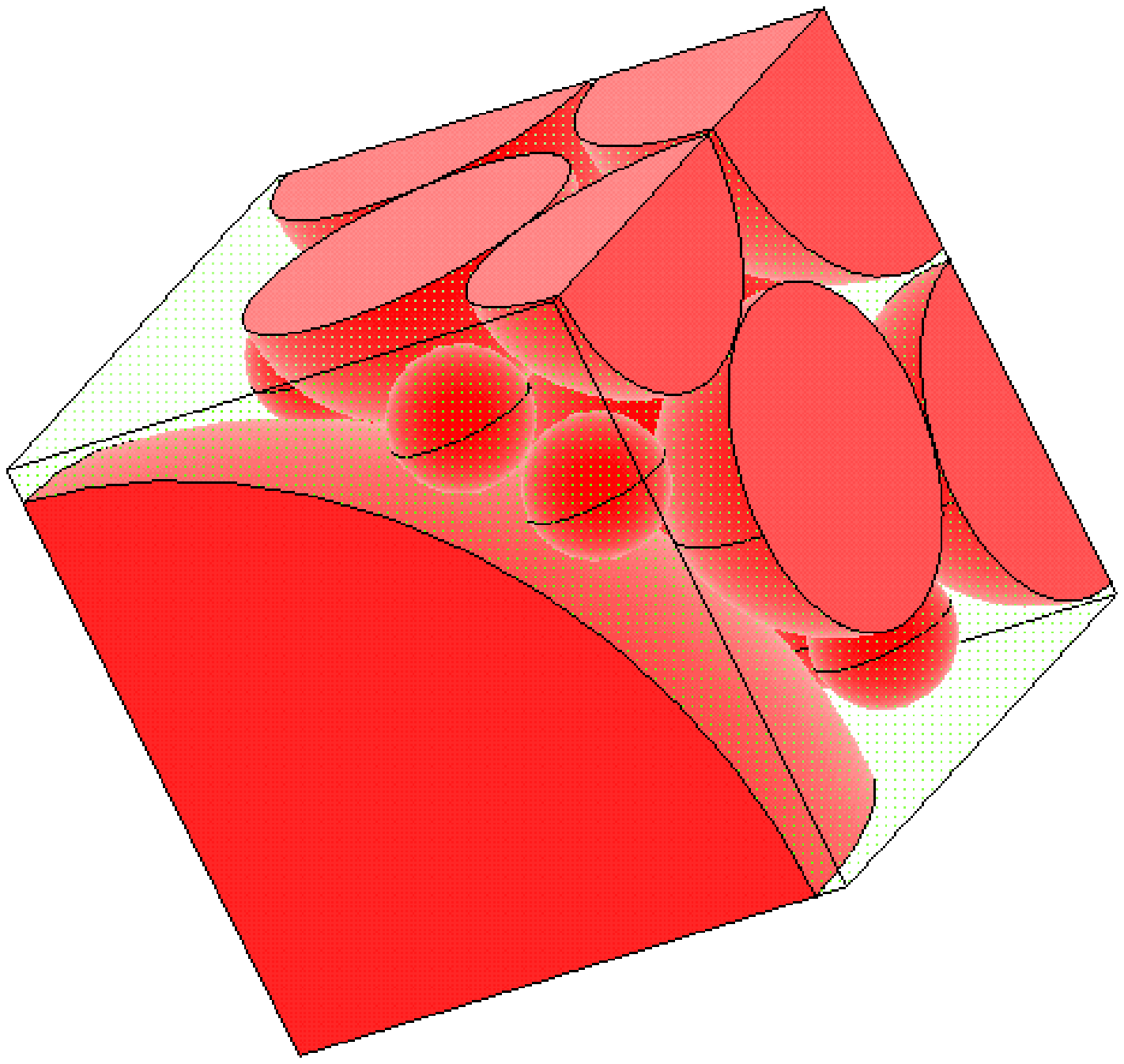}
                     \includegraphics{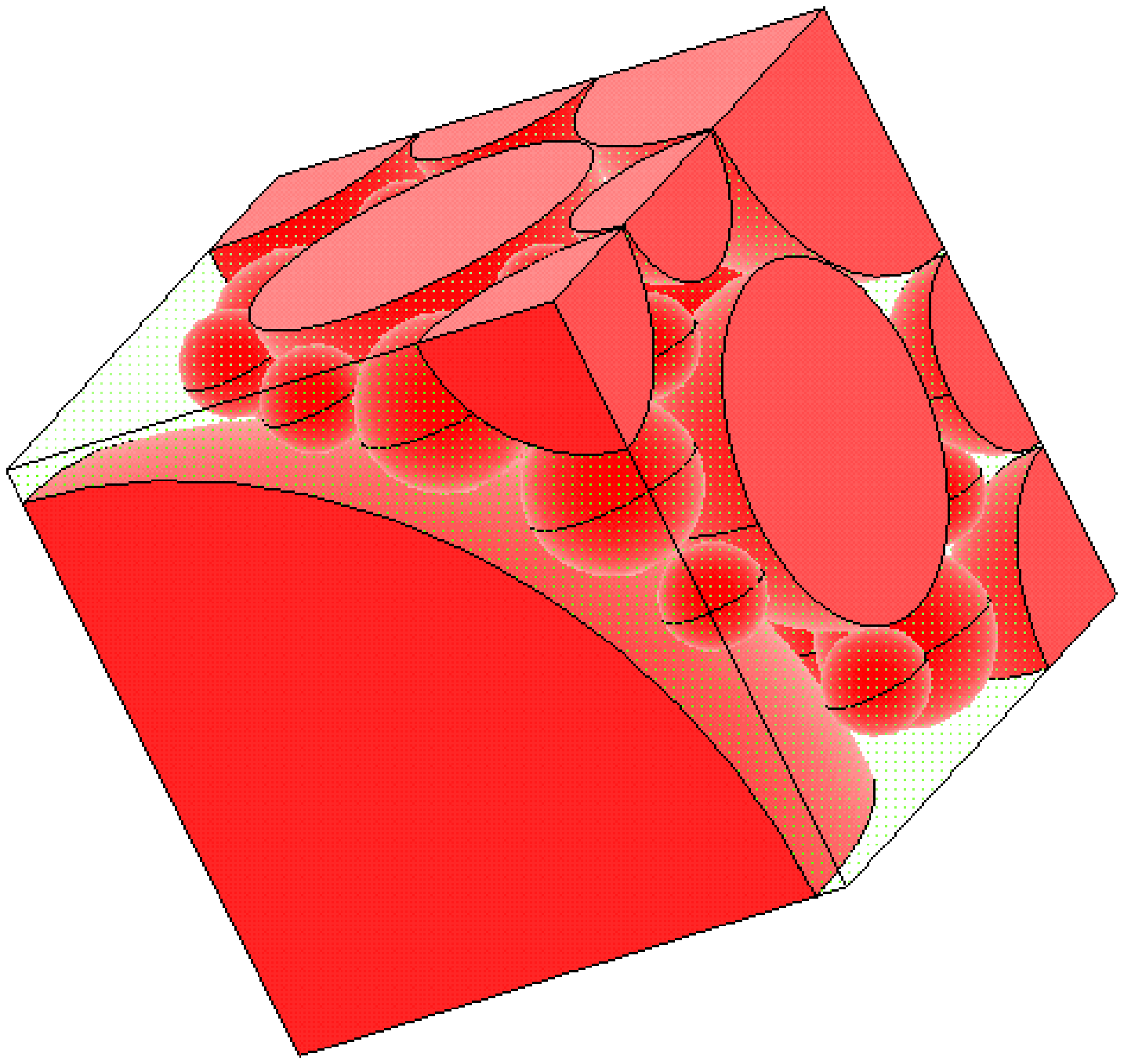}
                     \includegraphics{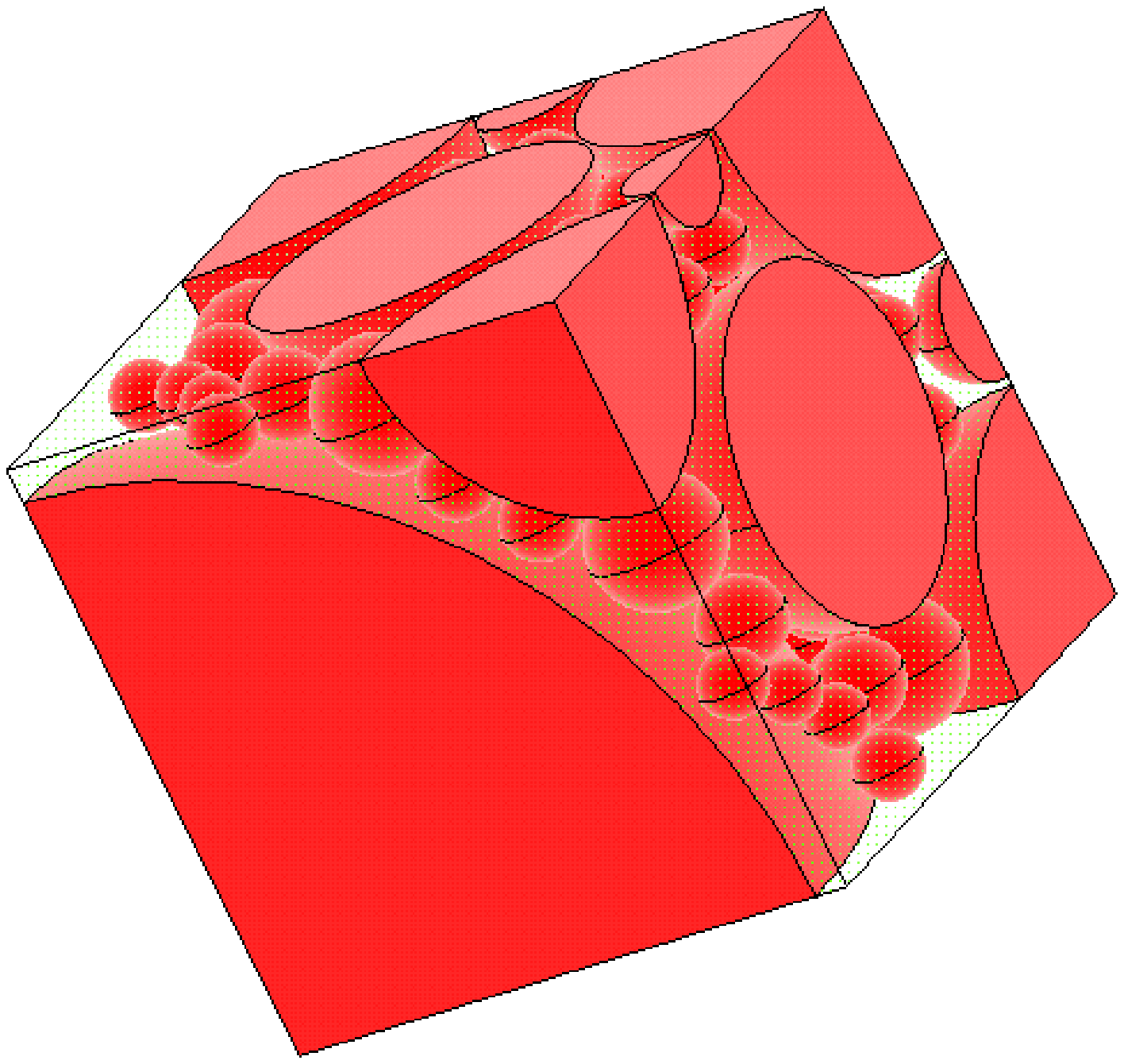}}\\
     (b) Three-dimensional RVES.
    \end{center}
     \caption{Two- and three-dimensional RVEs containing 70\%, 75\% and 
              80\% particles by volume. $f_p$ is the volume fraction of 
              particles.}
     \label{fig:2D3DFEM}
  \end{figure}

  To obtain a more direct estimate of the effectiveness of the two-dimensional 
  FEM approach, a second set of finite element simulations was performed on 
  two- and three-dimensional RVEs containing particle volume fractions of 
  0.70, 0.75, and 0.80.  The two-dimensional RVEs are shown in 
  Figure~\ref{fig:2D3DFEM}(a) and the three-dimensional RVEs are shown 
  in Figure~\ref{fig:2D3DFEM}(b).  The Young's modulus used for the particles 
  was 15,300 MPa and the Poisson's ratio was 0.32.  The Young's modulus of 
  the binder was 0.7 MPa and the Poisson's ratio was 0.49.  These values 
  correspond to the moduli of HMX and the binder of PBX 9501 at room 
  temperature and low strain rate~\cite{Zaug98,Wetzel99}.  

  For the two-dimensional simulations, each RVE was discretized using 
  six-noded triangles and approximately 60,000 nodes.  Displacement boundary 
  conditions were applied and the average stresses and strain were used to 
  calculate two-dimensional effective Young's moduli and Poisson's 
  ratios.  The two-dimensional moduli were then converted into 
  three-dimensional moduli using equations~\ref{eq:2Dto3D}.  The 
  three-dimensional RVEs were discretized using 10-noded tetrahedral 
  elements with a total of around 100,000 nodes.  A uniform displacement 
  was applied perpendicular to one of the faces of the RVE while the 
  remaining faces were constrained to displace in a periodic manner.  The 
  three-dimensional Young's modulus and Poisson's ratio were directly 
  calculated from the average stresses and strains for these three-dimensional 
  models.

\begin{figure}
    \begin{center}
     \scalebox{0.45}{\includegraphics{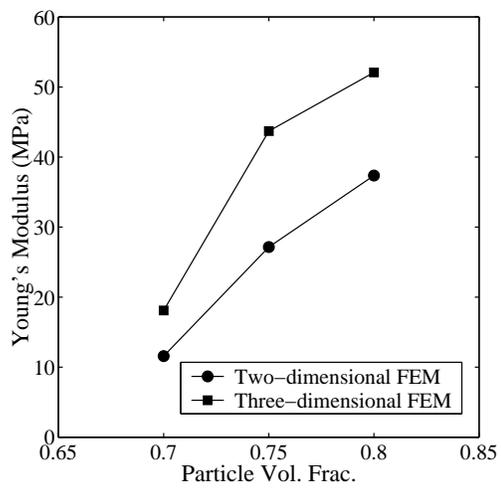}\hspace{24pt}
                     \includegraphics{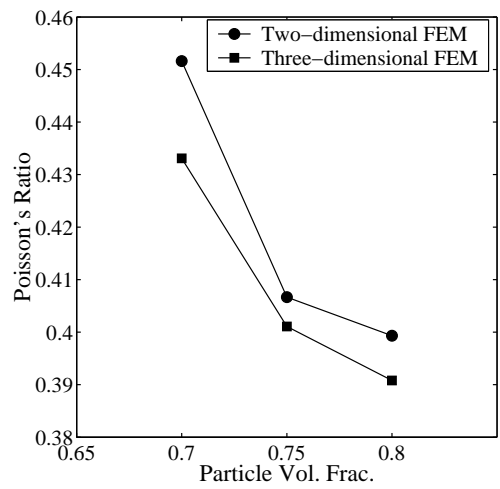}}\\
     (a)  Young's Modulus. \hspace{1.1in}(b) Poisson's Ratio.
    \end{center}
     \caption{Two- and three-dimensional estimates of effective Young's modulus
	      and Poisson's ratio.}
     \label{fig:2D3DEnu}
  \end{figure}

  Figure~\ref{fig:2D3DEnu}(a) shows the predicted effective Young's moduli 
  of the two- and three-dimensional RVEs shown in Figure~\ref{fig:2D3DFEM}.
  Though the three-dimensional models predict higher values of Young's 
  modulus than the two-dimensional models, the difference between the 
  two- and three-dimensional estimates of Young's modulus is small relative 
  to the modulus contrast between the particles and the binder.  
  Figure~\ref{fig:2D3DEnu}(b) shows the effective Poisson's ratio for the 
  two- and three-dimensional finite element models.  The three-dimensional 
  models predict lower values of Poisson's ratio than the 
  two-dimensional models.  Both two- and three-dimensional models predict 
  a sharp decrease in Poisson's ratio with increasing particle volume fraction.
  These results show that the two-dimensional FEM approach can be used to
  obtain acceptable estimates of effective elastic properties of random
  particulate composites and hence as a substitute for detailed 
  three-dimensional calculations.

  \section{Modeling polymer bonded explosives}
\begin{figure}
     \begin{center}
	\scalebox{0.7}{\includegraphics{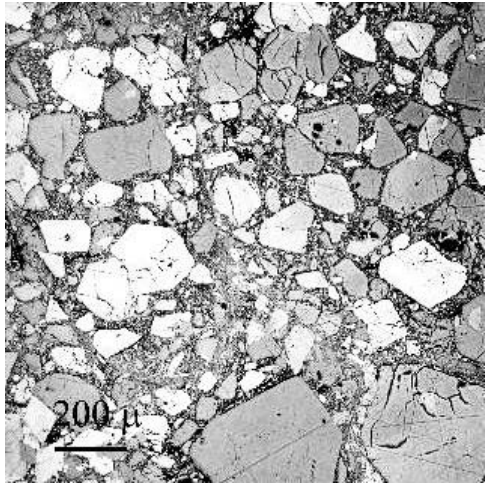}}
     \end{center}
 	\caption{Microstructure of PBX 9501 (adapted from ~\cite{Skid98}).
        The largest particles are around 300 microns in size.}
        \label{fig:skidmore}
  \end{figure}

  A micrograph of the polymer bonded explosive PBX 9501~\cite{Skid98} is shown 
  in Figure~\ref{fig:skidmore}.  The HMX particles are shown to be irregularly 
  shaped and present in a large number of sizes.  Two length scales can be 
  identified from the micrograph.  The first is the scale of the larger 
  particles that occupy most of the volume.  The second scale is that of the 
  particles filling the interstitial spaces between the larger 
  particles.  Because of these two different length scales, it is extremely 
  difficult to use a digital image at a single scale to generate 
  two-dimensional microstructures of PBX 9501 or other PBXs.  

  Most micromechanical calculations for PBXs have been carried out using 
  subgrid models that use simplified models of the microstructure 
  (for example, spherical grains coated with binder or spherical voids in an 
  effective PBX material~\cite{Massoni98}).  Closed form solutions from these 
  simple models have been used to provide properties for macroscopic 
  simulations. More detailed calculations have used microstructures containing 
  ordered arrays of circles or polygons in two or three dimensions to model 
  PBXs~\cite{Baer98, Hubner99}. These models do not reflect the microstructure 
  of PBXs and hence have limited use for predicting thermoelastic 
  properties.  Better two-dimensional approximations of the microstructure 
  have been constructed from digital images of the material and used by 
  Benson and Conley~\cite{Benson99} to study some aspects of the 
  micromechanics of PBXs.  However, such microstructures are extremely 
  difficult to generate and require state-of-the-art image processing 
  techniques and excellent images to accurately capture details of the 
  geometry of PBXs.  More recently Baer~\cite{Baer01} has used a combination 
  of Monte Carlo and molecular dynamics techniques to generate 
  three-dimensional microstructures containing spheres and oriented cubes 
  that appear to represent PBX microstructures well.  However, the generation 
  of a single realization of these microstructures is very time consuming and 
  often leads to a maximum packing fraction of about 70\%.  Periodicity is 
  also extremely difficult to maintain in the RVEs generated by this method.  

  One of the goals of this work was to explore a fast, yet accurate,
  means of predicting the effective elastic properties of PBXs.  With this
  goal in mind, accurate representation of the microstructure of PBXs was
  foregone.  Instead, the explosive crystals were modeled using
  circles or squares.  

  Both manually and automatically generated two-dimensional RVEs of PBX 9501 
  are discussed in this section.  The effective elastic moduli of these RVEs 
  from finite element calculations are then compared with experimentally 
  determined moduli of PBX 9501.  As in the previous section, the particles 
  in the RVEs represent HMX and have a Young's modulus of 15,300 MPa and a 
  Poisson's ratio of 0.32~\cite{Zaug98} and the binder has a Young's 
  modulus of 0.7 MPa and a Poisson's ratio of 0.49~\cite{Wetzel99}.  The 
  effective properties computed from these component properties are 
  compared with the Young's modulus of 1,013 MPa and Poisson's ratio 
  of 0.35~\cite{Wetzel99} of PBX 9501.

  \subsection{Manually generated microstructures}
\begin{figure}
     \begin{center}
	\scalebox{0.65}{\includegraphics{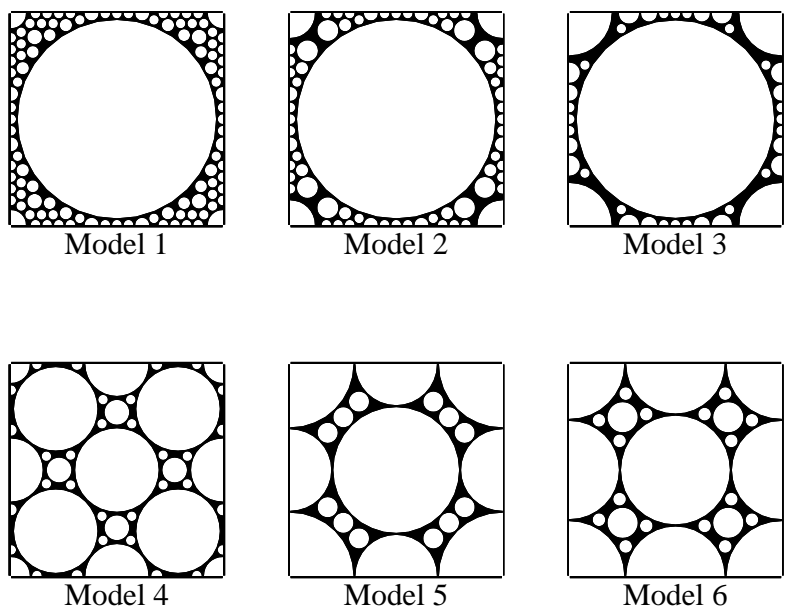}}
 	\caption{Manually generated microstructures containing $\sim$ 90\% 
                 by volume of circular particles.}
        \label{fig:pbxManual}
     \end{center}
  \end{figure}

  Figure~\ref{fig:pbxManual} shows six two-dimensional microstructures 
  representing PBX 9501.  Each RVE is filled with particles of various sizes 
  since PBXs are typically a mixture of coarse and fine grains with the finer 
  grains forming a filler between coarser grains.  The volume fractions of 
  particles in each of these models is 90$\pm$0.5\%.  All the models possess 
  square symmetry.  The RVEs have also been designed so that particle-particle 
  contact is avoided.
  Finite element simulations were performed on these RVEs using six-noded 
  triangular elements such that the circular particles were discretized 
  accurately.  The approach discussed in Section~\ref{sec:feApproach} was 
  used to determine the effective Young's modulus and Poisson's ratios of 
  these RVEs.  

  Table~\ref{tab:pbxManualFEM} shows the effective Young's modulus and 
  Poisson's ratios of the six RVEs.
  The values of the effective Young's modulus range from 42 MPa to 192 MPa, 
  with a mean of 132 MPa and a standard deviation of 54 MPa. In comparison, 
  the Young's modulus of PBX 9501 is 1,013 MPa; around 500\% to 800\% of that 
  predicted using the six RVEs.   Among the six RVEs, the Young's modulus 
  appears to be higher for the models with lower amounts of binder along the 
  edges of the RVE.  Models 1 through 3 have a single large particle and many 
  smaller particles and show approximately the same effective behavior.  Model 
  4, with a smaller ratio between the radius of the largest and the smallest 
  particles, has a lower effective Young's modulus than the mean value.  

  The effective Poisson's ratios of the six RVEs range from 0.25 to 0.34, with 
  a mean of 0.33.  The mean Poisson's ratio of the RVEs is not much different 
  from the measured value for PBX 9501.  For models 1, 2, and 3, the effective 
  Poisson's ratio is within 10\% of that of PBX 9501.  However, the Poisson's 
  ratio of model 4 is 25\% higher and those of models 5 and 6 are 20\% and 
  28\% lower than the Poisson's ratio of PBX 9501, respectively.  In 
  comparison, the two-dimensional Poisson's ratios of models 4, 5, and 6 are 
  0.79, 0.39, and 0.33, respectively, whereas that of PBX 9501 is 0.54.
  Since these two-dimensional values reflect the actual strains obtained from 
  the two-dimensional finite element calculations, it is clear that when
  compared to PBX 9501, model 4 is too compliant while models 5 and 6 are too 
  stiff.  These observations are also attributable to the amount of binder 
  along the edge of a RVE.
 
  On average, the FEM estimate of Young's modulus is around 13\% of the 
  experimentally determined value for PBX 9501.  The reason for this 
  difference could be that preferential stress paths (stress-bridging) in 
  the microstructure of PBX 9501 lead to increased stiffness.  Results shown 
  in Figure~\ref{fig:2D3DEnu} indicate that the difference between estimates 
  from two- and three-dimensional models is not large.  Hence, it is unlikely 
  that the considerably lower stiffness of the six model RVEs compared to
  PBX 9501 is due to the use of two-dimensional models.

\begin{figure}
     \begin{center}
     \scalebox{0.48}{\includegraphics{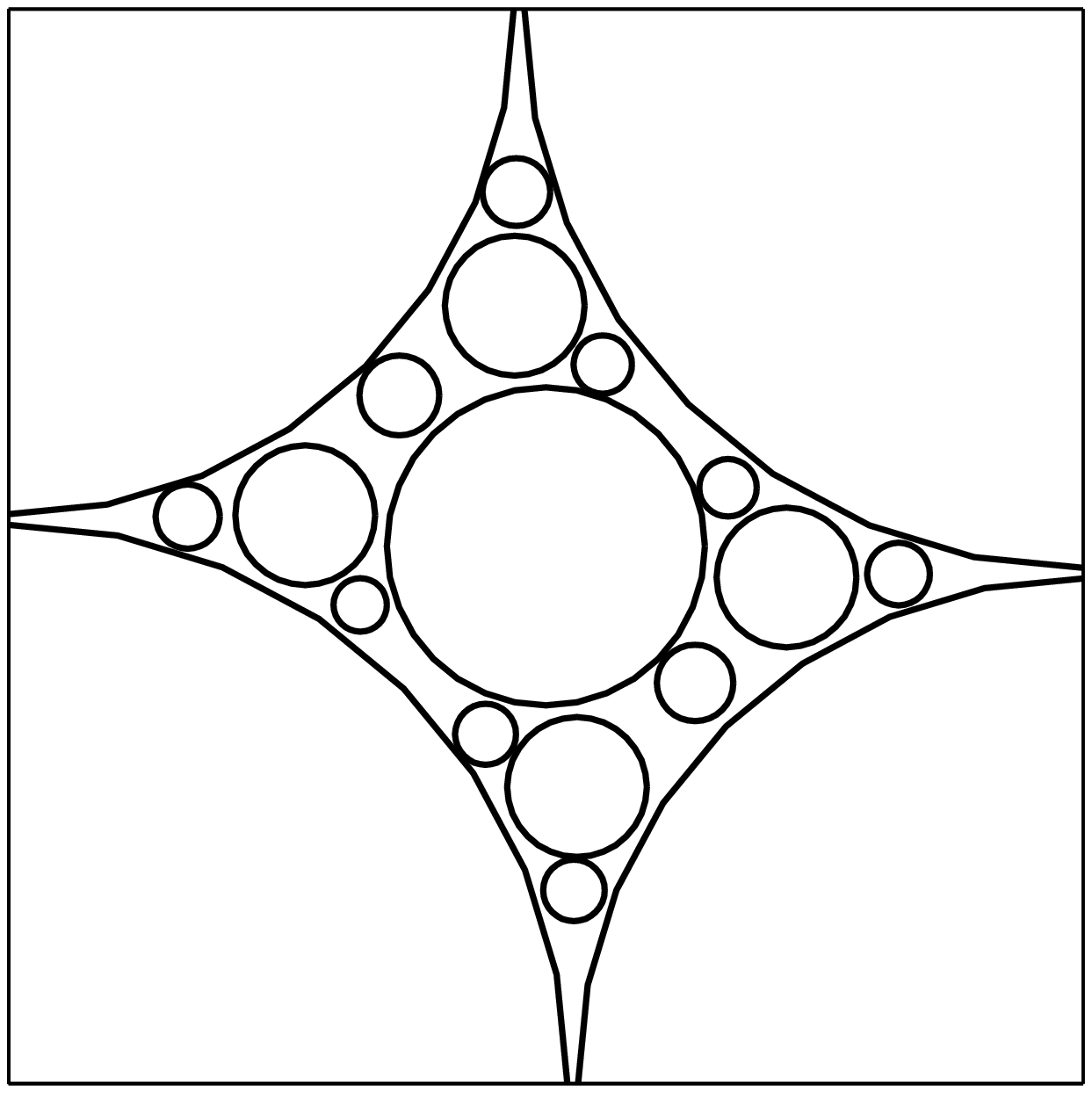}
                     \includegraphics{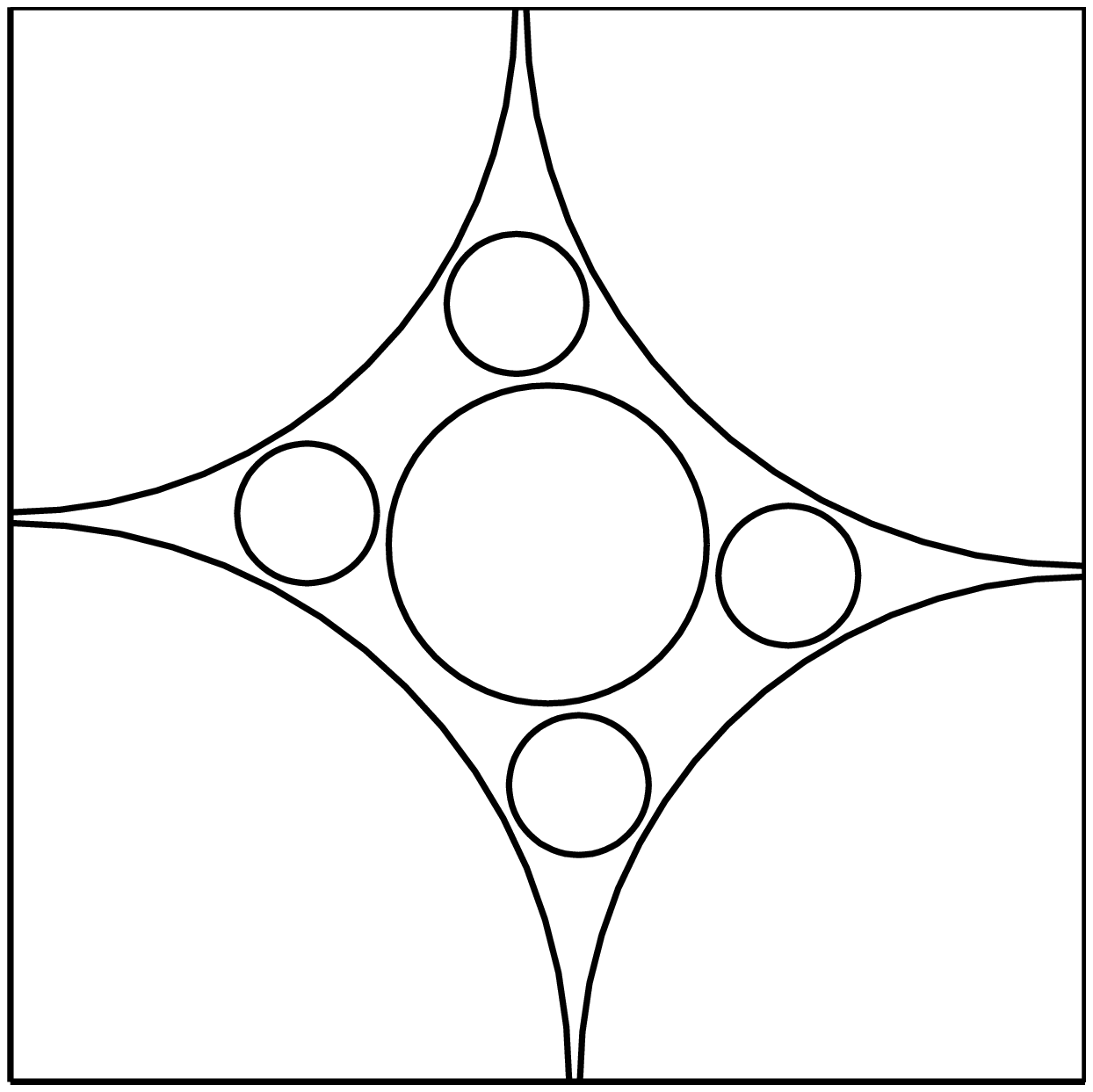}} \\
     (a) 92\% particles.\hspace{1.2in}(b) Model 6 - 90\% particles.\\ 
     \vspace{12pt}
     \scalebox{0.30}{\includegraphics{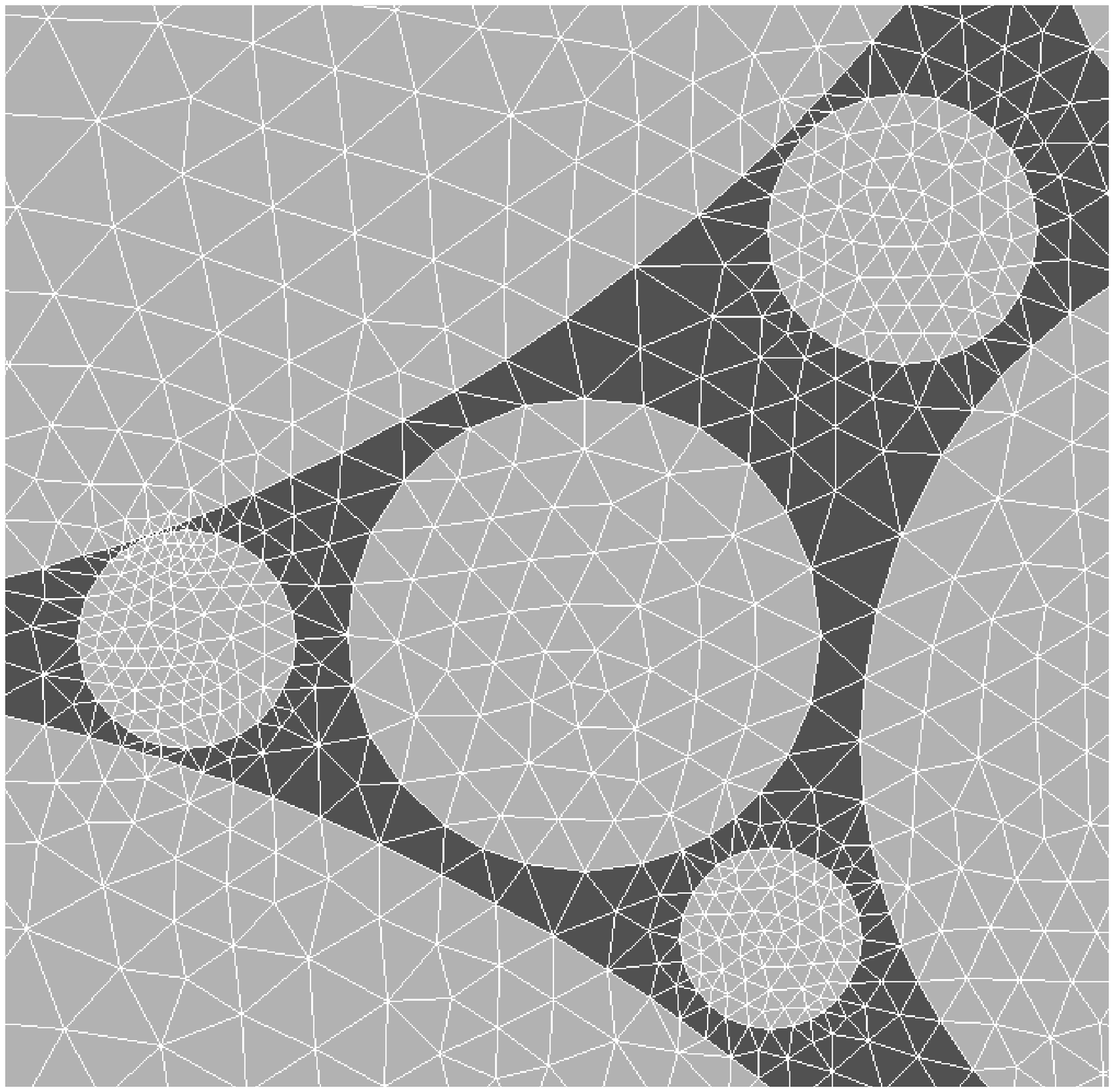}\hspace{36pt}
                     \includegraphics{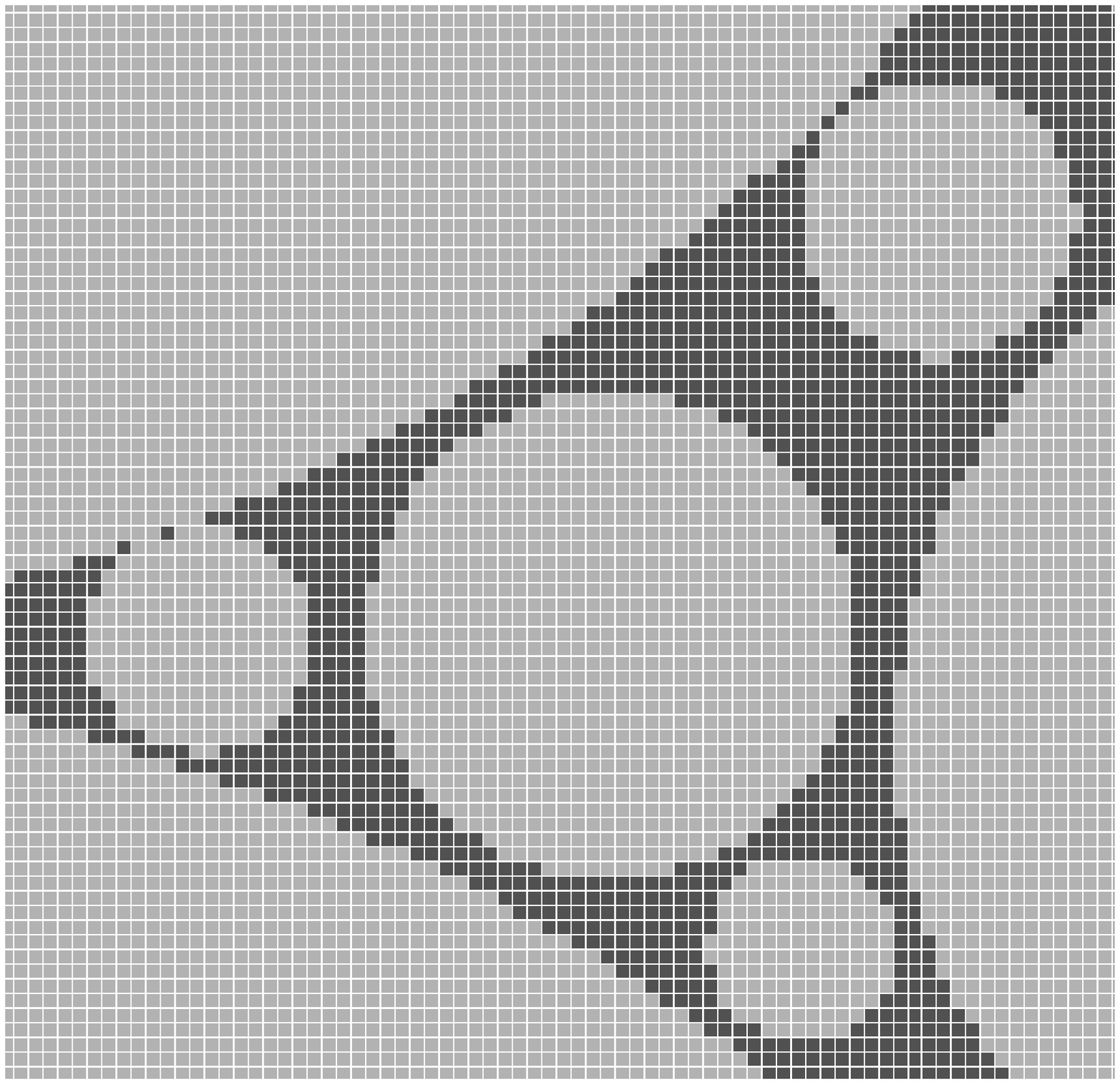}}\\
     (c) Six-noded Triangles.\hspace{1.2in}(d) Four-noded Squares.
     \end{center}
 	\caption{Microstructure containing 92\% particles modeled with 
                 six-noded triangles and four-noded squares.}
        \label{fig:pbxManual92}
  \end{figure}

  It is possible that the Young's modulus is underestimated by the six RVEs 
  because the volume fraction of particles in each of these microstructures 
  is lower than 92\%.  Figure~\ref{fig:pbxManual92}(a) shows a model 
  containing 92\% particles by volume that is a modified version of model 6 
  in Figure~\ref{fig:pbxManual}.  The effective Young's modulus of the RVE 
  containing 92\% particles is 218 MPa and the effective Poisson's ratio 
  is 0.28.  In comparison, model 6 contains around 90\% particles by volume 
  and has an estimated Young's modulus of 192 MPa and a Poisson's ratio of 
  0.25.  Though there is some increase in the Young's modulus due to increase 
  in particle volume fraction, the value is still around 20\% that of 
  PBX 9501.  Hence, the volume fraction cannot be the only factor leading 
  to underestimation of the Young's modulus by the manually generated 
  microstructures.  

  If a 256$\times$256 square grid is overlaid on top of the model shown in 
  Figure~\ref{fig:pbxManual92}(a) and cells in the grid are assigned HMX 
  properties if they contain more than 50\% particle and binder properties 
  otherwise (referred to as the "square grid overlay method"), the new 
  microstructure shown in Figure~\ref{fig:pbxManual92}(d) is obtained.  The 
  RVE in Figure~\ref{fig:pbxManual92}(d) was modeled using 256$\times$256 
  four-noded square finite elements.  The modified model predicts a Young's 
  modulus of 800 MPa and a Poisson's ratio of 0.14.  The modulus is now 
  closer to that of PBX 9501 but the Poisson's ratio is much lower.  

  The stiffer response of the model shown in Figure~\ref{fig:pbxManual92}(d) 
  is partially due to increased stiffness of four-noded finite elements.  To 
  eliminate the possibility that the stiffer response of the modified model 
  is due to element locking caused by the nearly incompressible binder, the 
  binder was also modeled as a two-parameter Mooney-Rivlin rubber instead of 
  a linear elastic material.  An incremental analysis using the Mooney-Rivlin 
  material model yielded values of Young's modulus and Poisson's ratio that 
  were within 5\% of those for the linear elastic models.  This result implies 
  that the difference between the effective properties of the RVEs in 
  Figures~\ref{fig:pbxManual92}(c) and~\ref{fig:pbxManual92}(d) is primarily 
  due to the slight difference in microstructure.  The same behavior has 
  also been observed for all the models shown in Figure~\ref{fig:pbxManual}.

  The model in Figure~\ref{fig:pbxManual92}(d) has a Young's modulus that is 
  four times higher than that of the model in 
  Figure~\ref{fig:pbxManual92}(c).  This result suggests that the square grid 
  overlay method can lead to the automatic incorporation of preferential 
  stress paths into RVEs containing high volume fractions of particles.  In 
  addition, it is easier to generate a square mesh and use the square grid 
  overlay method than to model circular particles using triangular 
  elements.  Hence, the square grid overlay method is used to discretize 
  the randomly generated microstructures based on the actual particle size 
  distribution of PBX 9501 discussed in the next section.

  The manually generated microstructures were based on a qualitative 
  assessment of the particle size distribution of PBX 9501 without regard
  for the actual size distribution.  In addition, particles in the RVEs
  were not allowed to come into direct contact leading to a gross
  underestimation of the effective Young's modulus.  The first of these
  issues is addressed in the following section where the actual particle
  size distribution of PBX 9501 is used to generate model microstructures.
  The particle contact problem is addressed by approximating the model
  microstructures using the square grid overlay method.
  
  \subsection{Randomly generated microstructures}
  The preferred method for generating close packed microstructures from a set 
  of particles is to use Monte Carlo based molecular dynamics 
  techniques~\cite{Tanemura92} or Newtonian motion based 
  techniques~\cite{Stroeven99}.  In both methods, a distribution of particles 
  is allocated to the grid points of a rectangular lattice using a random 
  placement method.  Molecular dynamics simulations or Newtonian dynamics 
  calculations are then carried out on the system of particles to reach the 
  packing fraction that corresponds to equilibrium.  A weighted Voronoi 
  tessellation~\cite{Auren91} is then carried out on the particles with the 
  weights determined by the sizes of the particles.  The particles are next 
  moved towards the center of the packing volume while maintaining that they 
  remain inside their respective Voronoi cells.  The process is repeated until 
  all the particles are as tightly packed as possible.  Periodicity of the 
  particles at the boundaries is maintained by specifying extra particles at 
  the boundaries that move in and out of the volume.  This process, with some 
  modifications, has been the only efficient method of generating close packed 
  systems of particles in three dimensions.  However, it is difficult to get 
  packing fractions of more than 70-75\% when using spherical particles.  It 
  is also quite time consuming to generate a tight packing.
  
  In two dimensions, a faster method can be used for generating particle 
  distributions, that is random sequential packing.  The largest particles are 
  placed randomly in the volume followed by progressively smaller 
  particles.  If there is any overlap between a new particle and the existing 
  set, the new particle is moved to a new position.  If a particle cannot be 
  placed in the volume after a certain number of iterations, the next lower 
  sized particle is chosen and the process is continued until the required 
  volume fraction is achieved.  Though this method does not preserve the 
  particle size distributions as accurately as the Monte Carlo based molecular 
  dynamics methods, it is much faster and can be used to generate high packing 
  fractions in two dimensions without particle locking.  In three 
  dimensions, this method is highly inefficient and packing above 60\% is 
  extremely time consuming to achieve.

  \subsubsection{Circular particles - PBX 9501 dry blend}
  HMX particle size distributions of the dry blend of PBX 9501 have been 
  listed by Wetzel~\cite{Wetzel99} and Rae et al.~\cite{Rae02}.  The coarse 
  and fine particles are blended in a ratio of 1:3 by weight and compacted 
  to form PBX 9501.  Four microstructures based on the particle size 
  distribution of the dry blend are shown in Figure~\ref{fig:dryBlendPBX}. The 
  number of particles used for the four microstructures are 100, 200, 300, and 
  400.  The RVE widths are 0.65 mm, 0.94 mm, 1.13 mm and 1.325 mm, 
  respectively.  The particles occupy a volume fraction of about 85-86\%.  The 
  remaining volume is assumed to be occupied by a mixture of binder and fine 
  particles of HMX that are well separated in size from the smallest particles 
  shown in the RVEs.  This `dirty' binder contains around 36\% particles by 
  volume.

  Particles in the RVEs were assigned HMX properties and the elastic properties
  of the dirty binder were calculated using the differential effective medium 
 approximation~\cite{Markov00}.  The Young's modulus of the dirty binder used 
 in the calculations was 2.0816 MPa and the Poisson's ratio was 0.4813.  Two 
 sizes of square grids were used for overlaying on the RVEs: 256$\times$256 
 and 350$\times$350.  The square grid overlay method was applied to assign 
 materials to grid cells.  Each cell in the grid was modeled using a 
 four-noded finite element.  Table~\ref{tab:dryBlendPBX} shows the effective 
 Young's modulus and Poisson's ratio predicted for the four models of the dry 
 blend of PBX 9501. 

  The FEM calculations overestimate the effective Young's moduli of PBX 9501 
  by 200\% to 400\% if the 256$\times$256 grid is used.  If the 350$\times$350 
  grid is used the effective Young's moduli vary from 100\% to 300\% of that 
  of PBX 9501.  The effective Poisson's ratio is underestimated considerably 
  in both cases.  The stiffer response of these RVEs is partly due to the 
  creation of continuous stress paths across a RVE when a regular grid is 
  overlaid on the RVE and partly because four-noded elements are inherently 
  stiffer~\cite{Bathe97}.  It can also be seen that the effective Young's 
  modulus increases as the number of particles in the RVE increases.  This is 
  because the same grid has been used for the four microstructures, leading 
  to poorer resolution of the geometry with increasing particle density per 
  grid cell.  

  \subsubsection{Circular particles - pressed PBX 9501}
  Figure~\ref{fig:pressedPBX} shows four RVEs based on the particle size 
  distribution of pressed PBX 9501~\cite{Skid98}.  
  The pressing process leads to particle breakage and hence a larger volume 
  fraction of smaller sized particles.  Fewer larger sized particles remain 
  as is reflected in the generated microstructures containing 100, 200, 500, 
  and 1000 particles.  In this case, the RVE widths are 0.36 mm, 0.42 mm, 
  0.535 mm, and 0.68 mm, respectively.  Thus, the 1000 particle RVE for the 
  pressed piece has dimensions similar to the 100 particle RVE for the dry 
  blend.  The size of the RVE that can be adequately discretized is therefore 
  smaller for the pressed piece.  Each RVE was discretized into both 
  256$\times$256 and 350$\times$350 four-noded elements.  Each element was 
  assigned material properties according to the square grid overlay method.  

  The particles in the RVEs occupy volume fractions from 0.86 to 0.89 and 
  the target volume fraction of 0.92 is not attained in any of the RVEs.  A 
  dirty binder, whose properties were calculated using the differential 
  effective medium approach, was used in the effective stiffness 
  calculations.  The effective moduli of the dirty binder for the four 
  models are shown in Table~\ref{tab:pressedPBX}.  
  The effective Young's modulus and Poisson's ratio of the four RVEs from FEM 
  calculations are shown in Table~\ref{tab:pressedPBXCircle}.
  The 256$\times$256 element models overestimate the Young's modulus of PBX 
  9501 by factors increasing from 3 to 6 with increasing RVE size.  The 
  estimates from the 350$\times$350 element models are lower but still 2 to 
  5 times higher than the Young's modulus of PBX 9501.  The estimated 
  Poisson's ratios are quite low compared to that of PBX 9501.  

  The four models based on pressed PBX 9501 predict Young's moduli that are 
  1.5 to 2 times higher than values predicted by models based on the dry blend 
  of PBX 9501.  For the 100 and 200 particle pressed PBX 9501 models, the 
  single large particle contributes considerably to the stiffer response.  For 
  the 500 and 1000 particle pressed PBX 9501 models, errors in the 
  discretization of particle boundaries lead to additional stress bridging 
  paths and hence a stiffer response is obtained.  
  
  \subsubsection{Square particles - pressed PBX 9501}
  With an increase in particle volume fraction, there is an increase in the 
  number of particle-particle contacts in particulate composites.  If the 
  particles in a RVE of such a material are circular, triangular elements 
  cannot be used to discretize the RVE because of poor element shapes in 
  regions of contact.  Additionally, a rectangular grid cannot represent 
  the geometry of circular particles accurately.  An alternative is to use 
  square particles that are aligned with a rectangular grid to represent the 
  microstructure.

  The particles shown in the three microstructures in 
  Figure~\ref{fig:pressedPBXsq} are based on the size distribution of 
  pressed PBX 9501.  
  These distributions have been generated by placing square particles in a 
  random sequential manner in a square grid.  The smallest particles occupy a 
  single subcell of the grid.  Larger particles are chosen from the particle 
  size distribution so that they fit into an integer multiple of the grid 
  size.  

  In the three models shown in Figure~\ref{fig:pressedPBXsq}, the particle 
  size distribution for the pressed piece is truncated so that the 
  smallest ($\le$ 30 microns) particle sizes in the distribution are not used 
  in order that the grid size does not become too large.  The RVEs are filled 
  with particles to volume fractions of about 86-87\%.  The remaining volume 
  is assumed to be occupied by a dirty binder.  The number of particles in the 
  first model is 700 and the RVE width is around 3.6 mm.  The second model 
  contains 2,800 particles and the RVE width is about 5.3 mm, while the third 
  model contains 11,600 particles and has a width of 9 mm.  Note that these 
  RVEs are considerably larger than those used for the circular 
  particles.  Smaller RVEs are not used because of the difficulties associated 
  with fitting particles into integer multiples of subcell widths.  The 
  properties of the dirty binder for the three models are shown in 
  Table~\ref{tab:pressedPBXsq}.

  Finite element calculations were performed on the three models using regular 
  grids of 256$\times$256 square elements.  Table~\ref{tab:pressedPBXSquare} 
  shows the effective Young's modulus and Poisson's ratio obtained from the 
  three RVEs.  The FEM estimates of Young's modulus are 9 times higher than 
  the Young's modulus of PBX 9501.  The higher stiffness obtained from the 
  FEM calculations is partly because the large number of contacts between 
  particles lead essentially to a continuous particle phase containing 
  pockets of binder.  

  Micrographs of PBX 9501 show that there is considerable damage in the 
  particles, binder and the particle-binder interface~\cite{Rae02}.  Better 
  estimates of the effective properties can probably be obtained by 
  incorporating damage in the models of PBX 9501 discussed above.  However, 
  this issue is not explored in this work because the amount and type of 
  damage is difficult to quantify in polymer bonded explosives. 

  \section{Effective moduli of PBX 9501}
\begin{figure}
     \begin{center}
	\scalebox{0.13}{\includegraphics{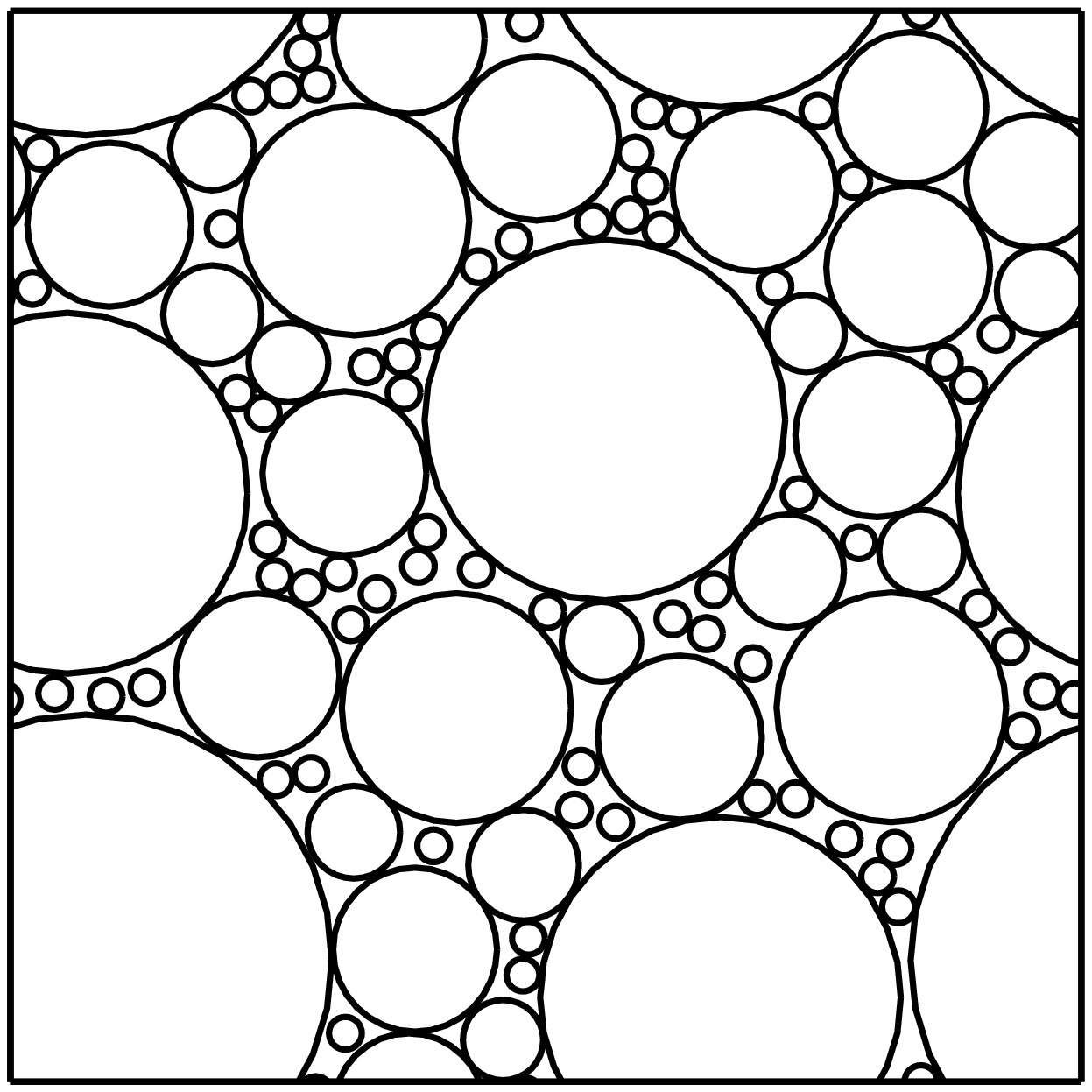}}\hfill
	\scalebox{0.20}{\includegraphics{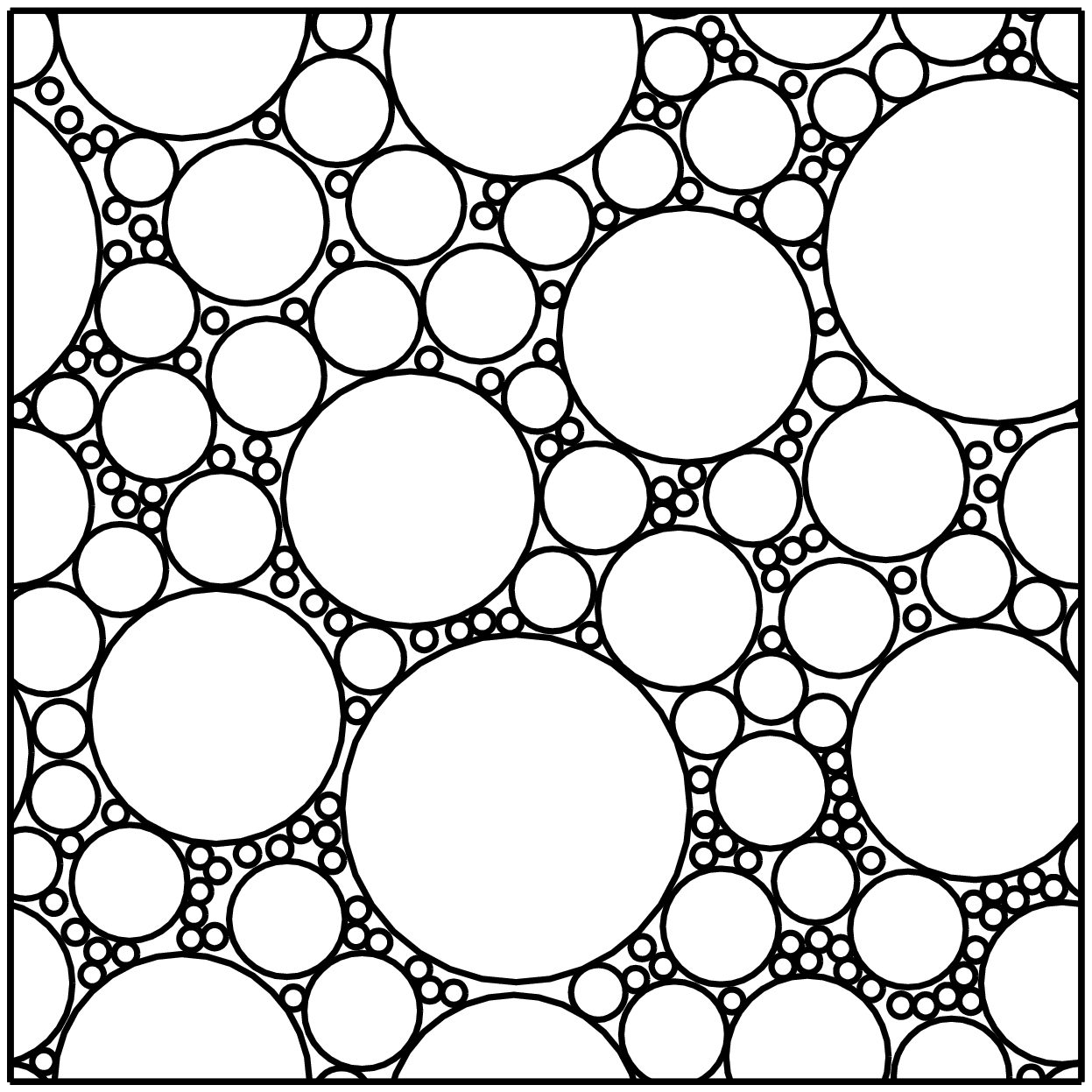}}\hfill
	\scalebox{0.24}{\includegraphics{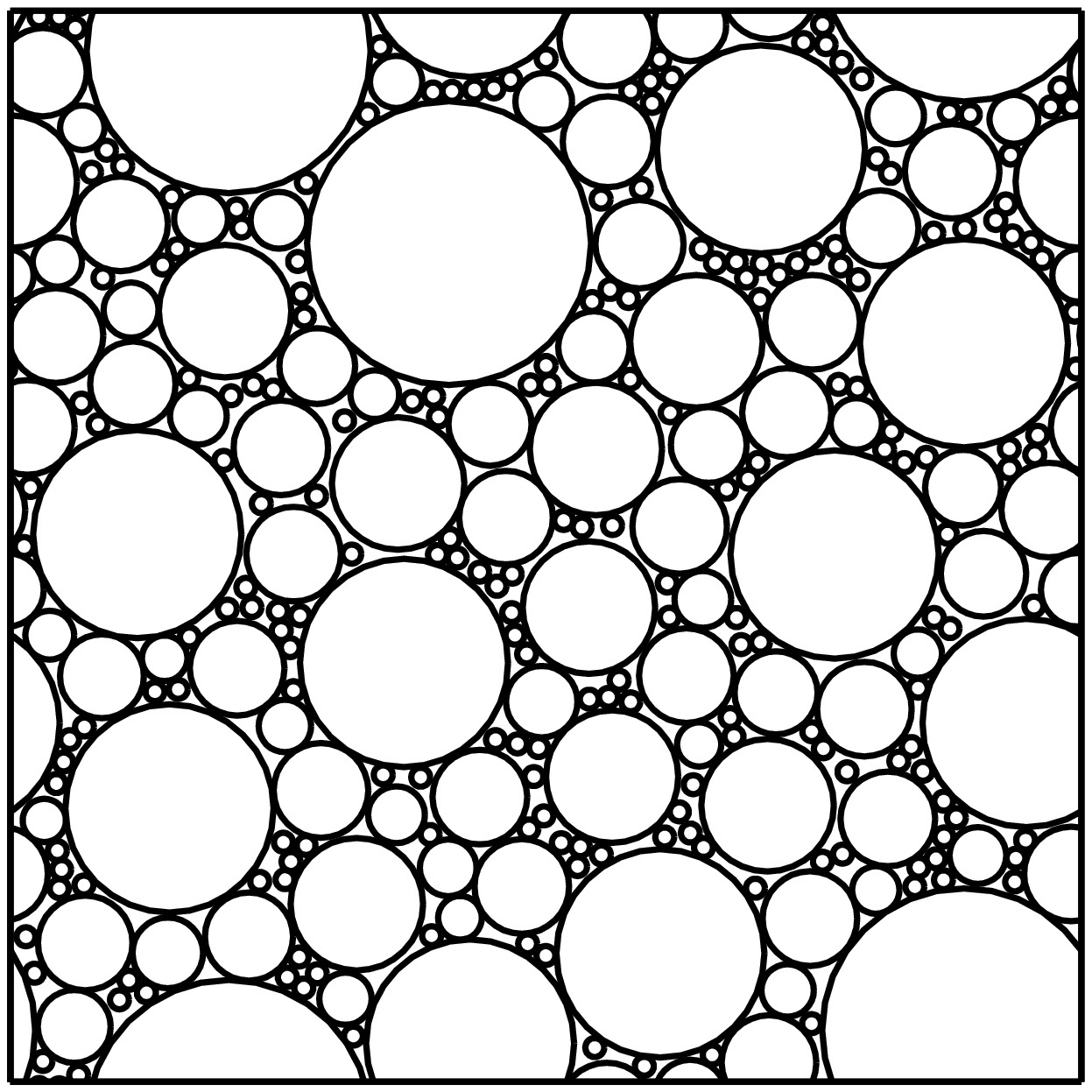}}\hfill
	\scalebox{0.28}{\includegraphics{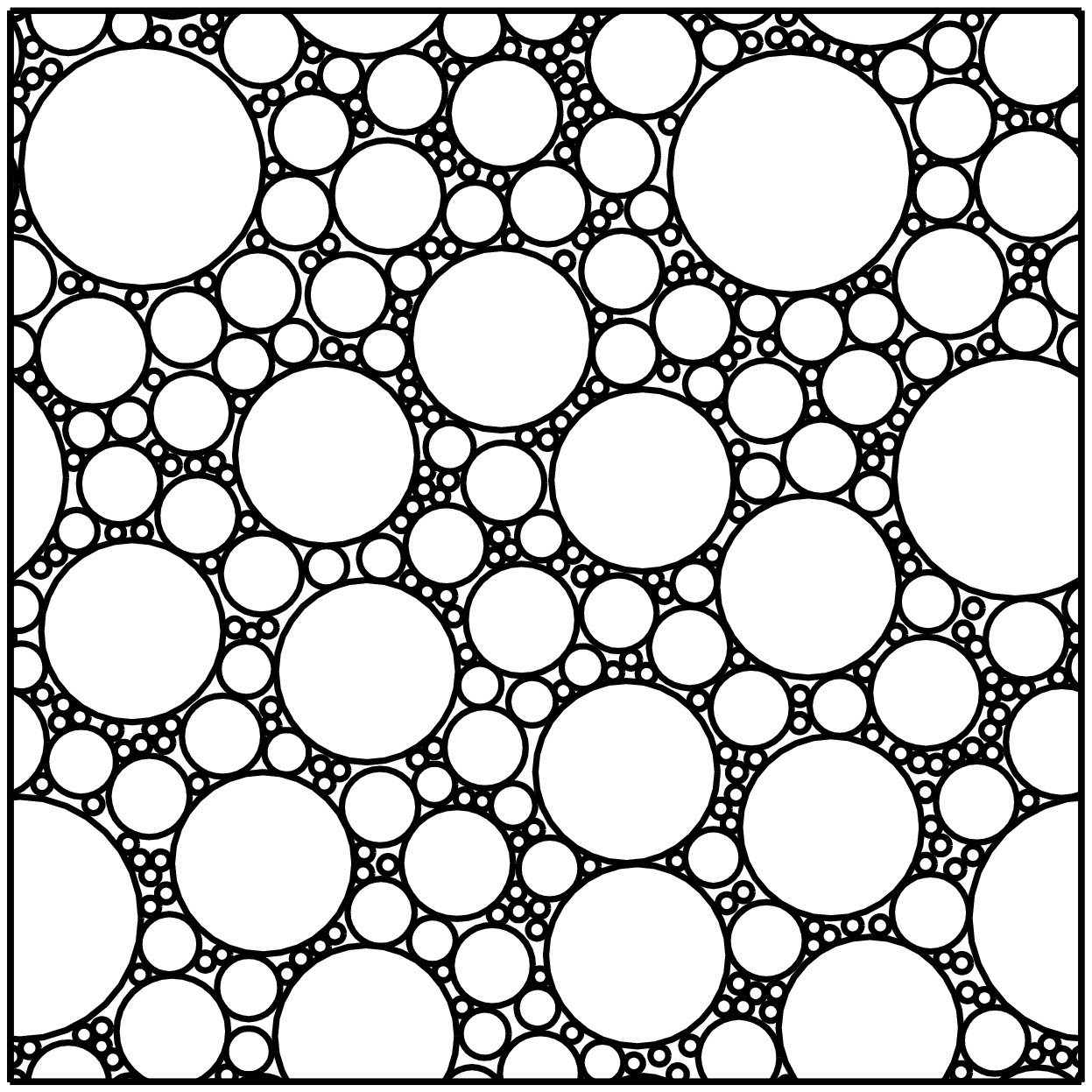}}\\
        0.65$\times$0.65 mm$^2$ \hspace{2pt}
        0.94$\times$0.94 mm$^2$ \hspace{32pt}
        1.13$\times$1.13 mm$^2$ \hfill
        1.33$\times$1.33 mm$^2$ \\
        100 Particles \hspace{26pt} 200 Particles \hspace{50pt}
        300 Particles \hfill 400 Particles
     \end{center}
 	\caption{Circular particle models based on the dry blend of PBX 9501.} 
        \label{fig:dryBlendPBX}
  \end{figure}

\begin{figure}
     \begin{center}
	\scalebox{0.16}{\includegraphics{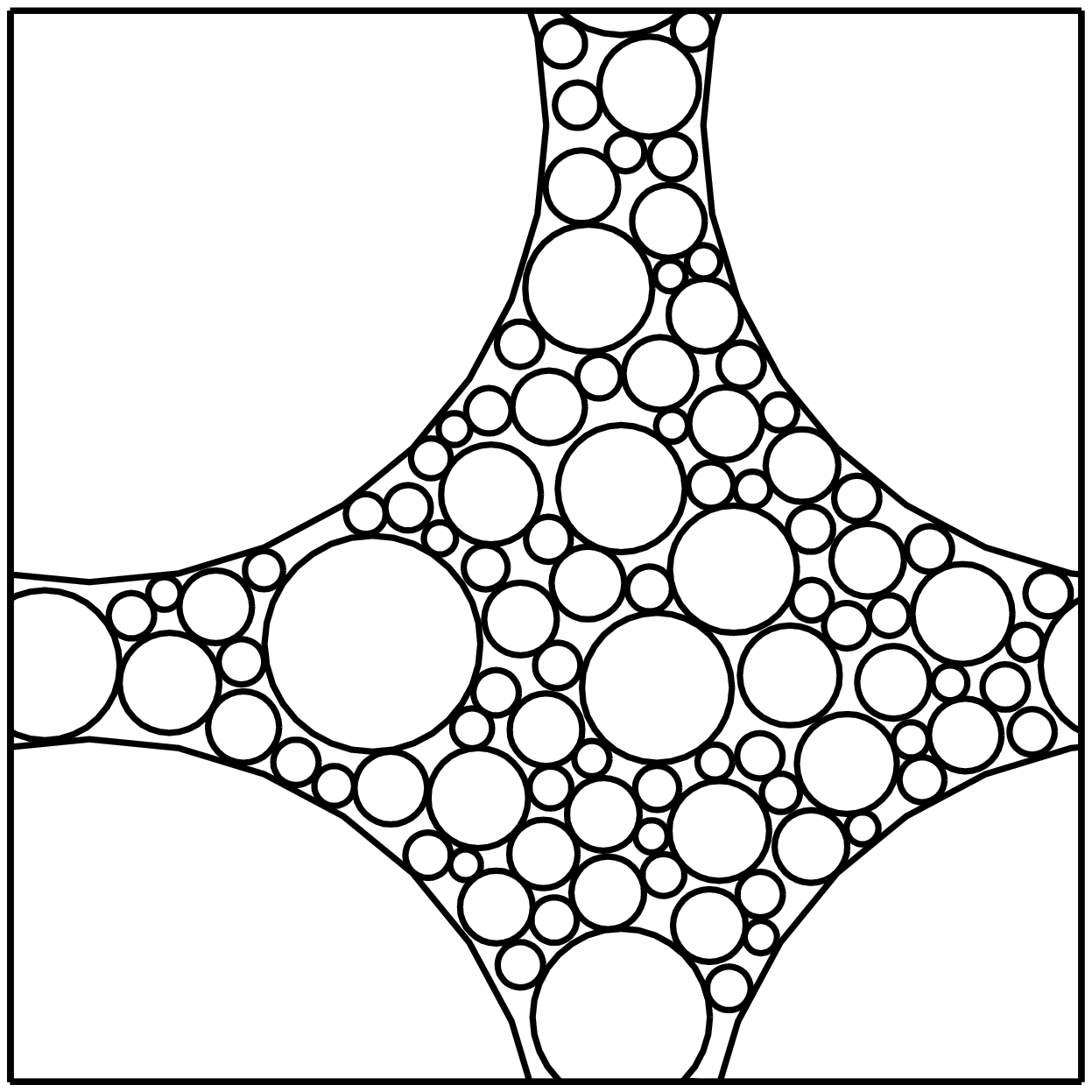}}\hfill
	\scalebox{0.19}{\includegraphics{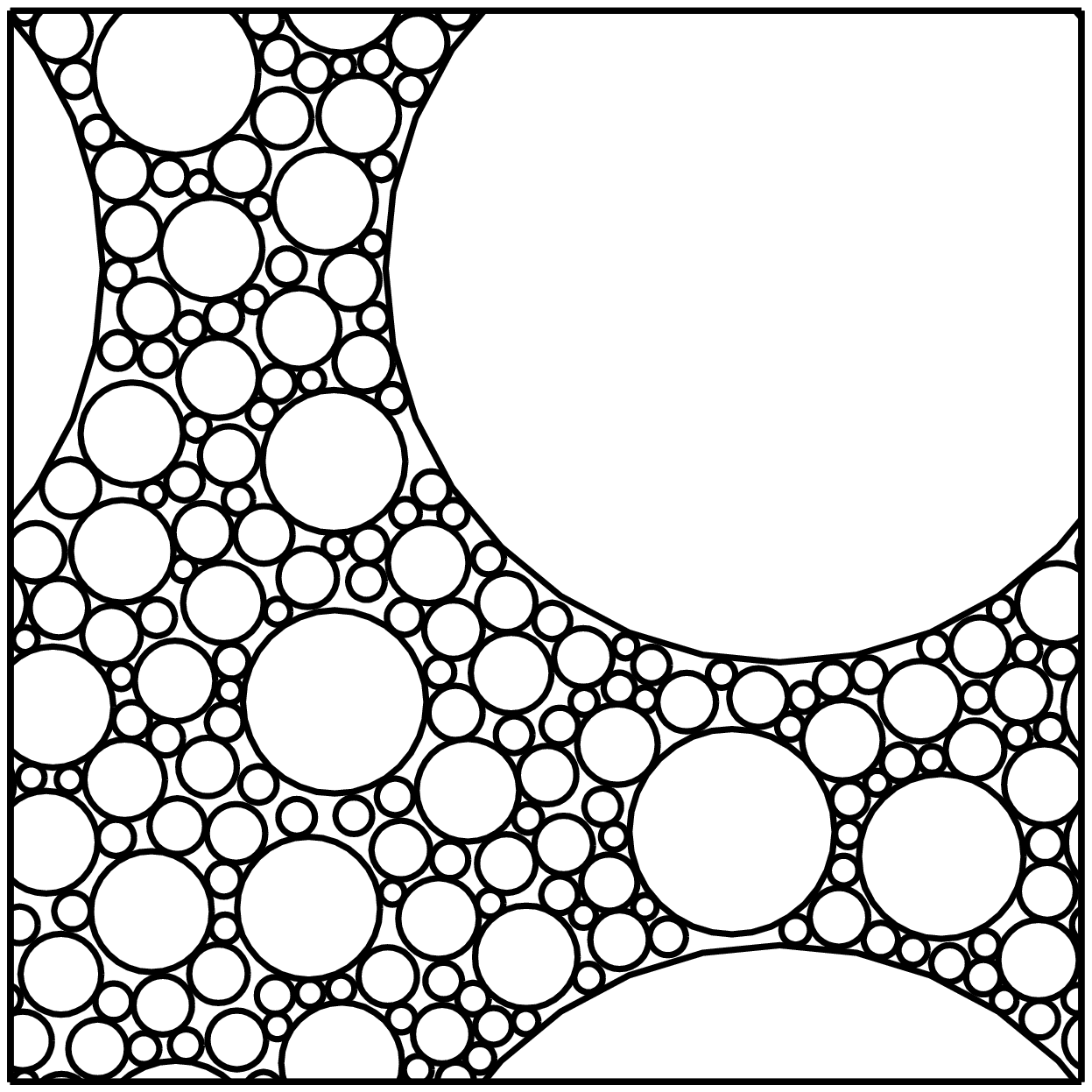}}\hfill
	\scalebox{0.24}{\includegraphics{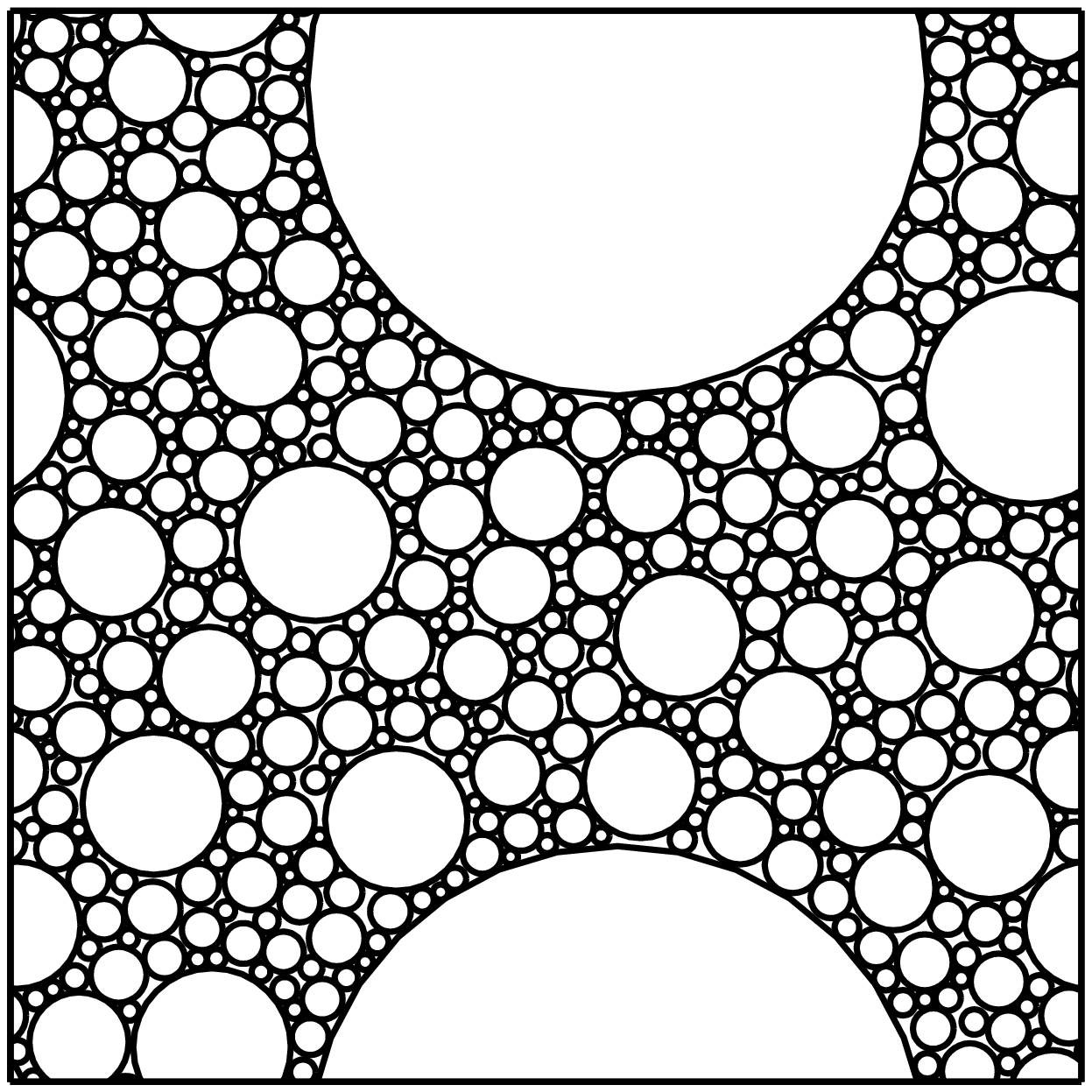}}\hfill
	\scalebox{0.30}{\includegraphics{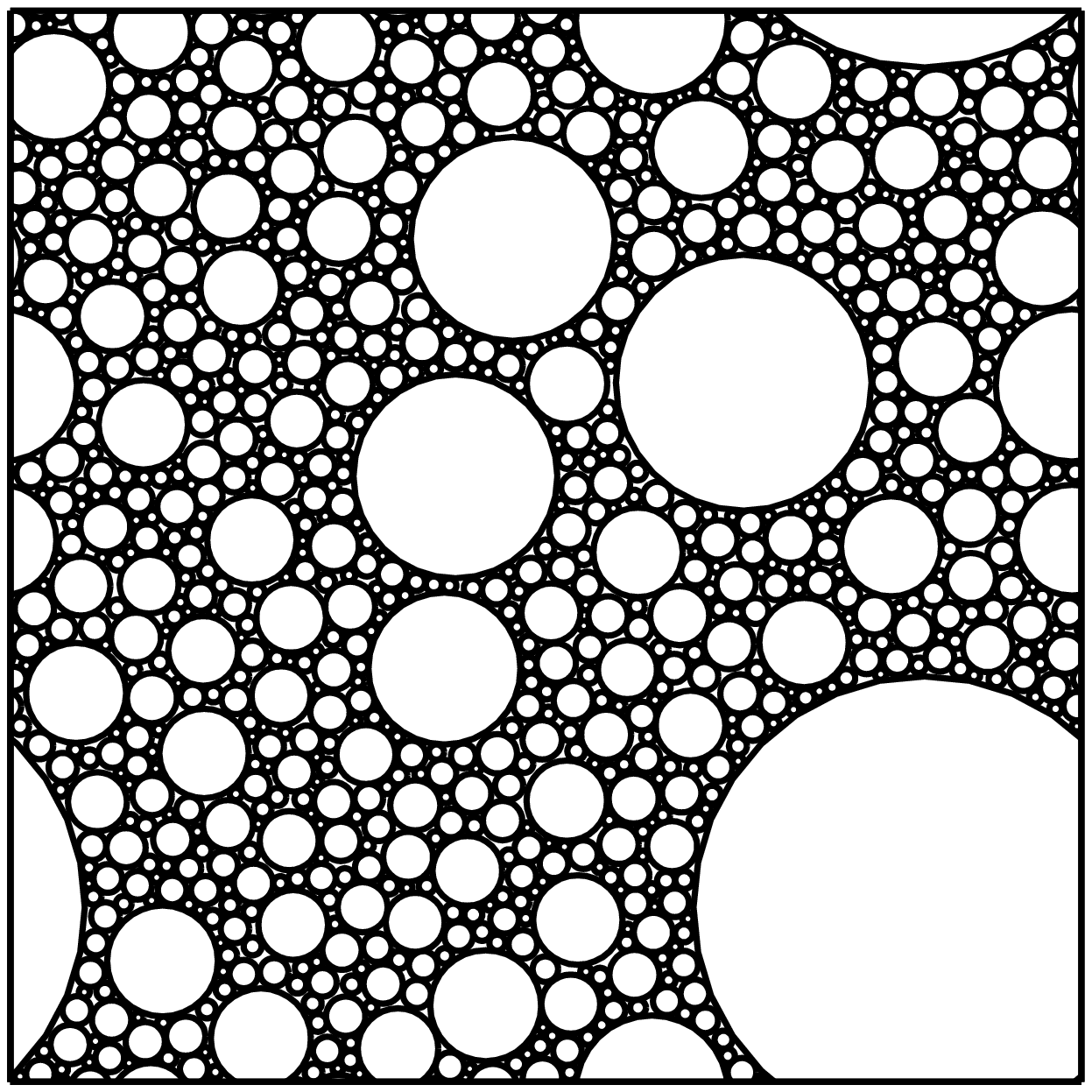}}\\
        0.36$\times$0.36 mm$^2$ \hspace{2pt}
        0.42$\times$0.42 mm$^2$ \hspace{30pt}
        0.54$\times$0.54 mm$^2$ \hfill
        0.68$\times$0.68 mm$^2$ \\
        100 Particles \hspace{27pt} 200 Particles \hspace{50pt}
        500 Particles \hfill 1000 Particles
             \end{center}
 	\caption{Circular particle models based on the 
		 particle size distribution of pressed PBX 9501.} 
        \label{fig:pressedPBX}
  \end{figure}

\begin{figure}
     \begin{center}
	\scalebox{0.16}{\includegraphics{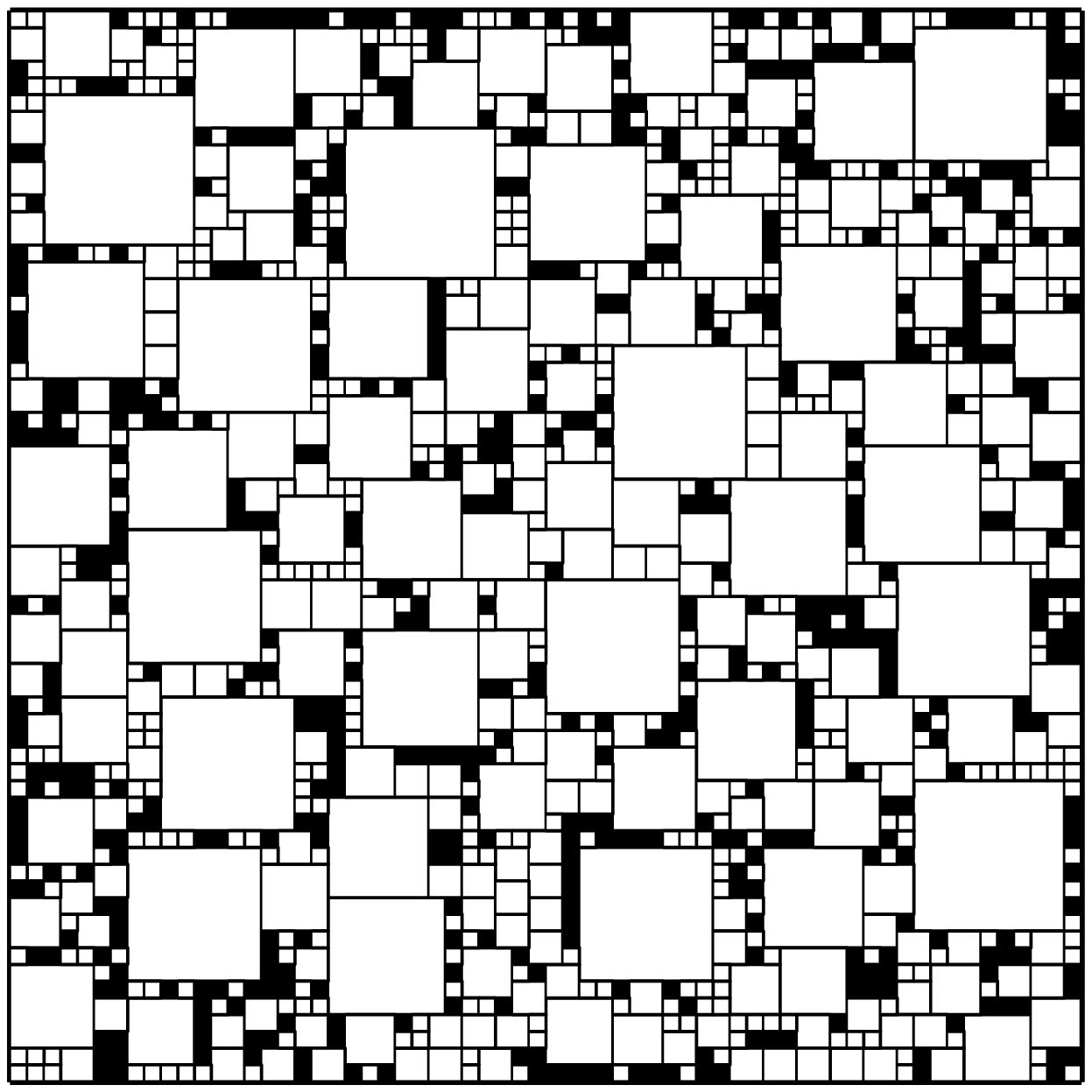}}\hfill
	\scalebox{0.24}{\includegraphics{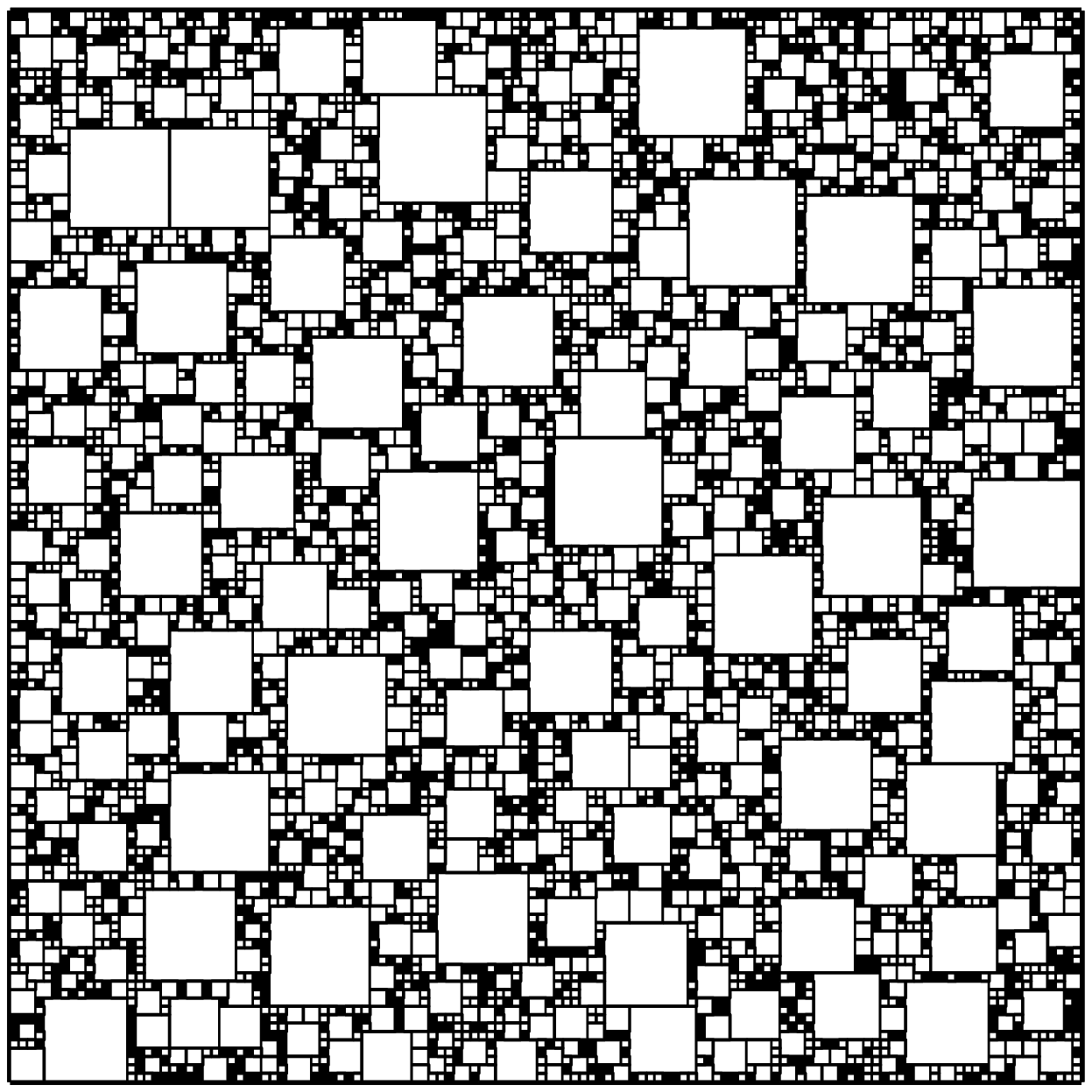}}\hfill
	\scalebox{0.40}{\includegraphics{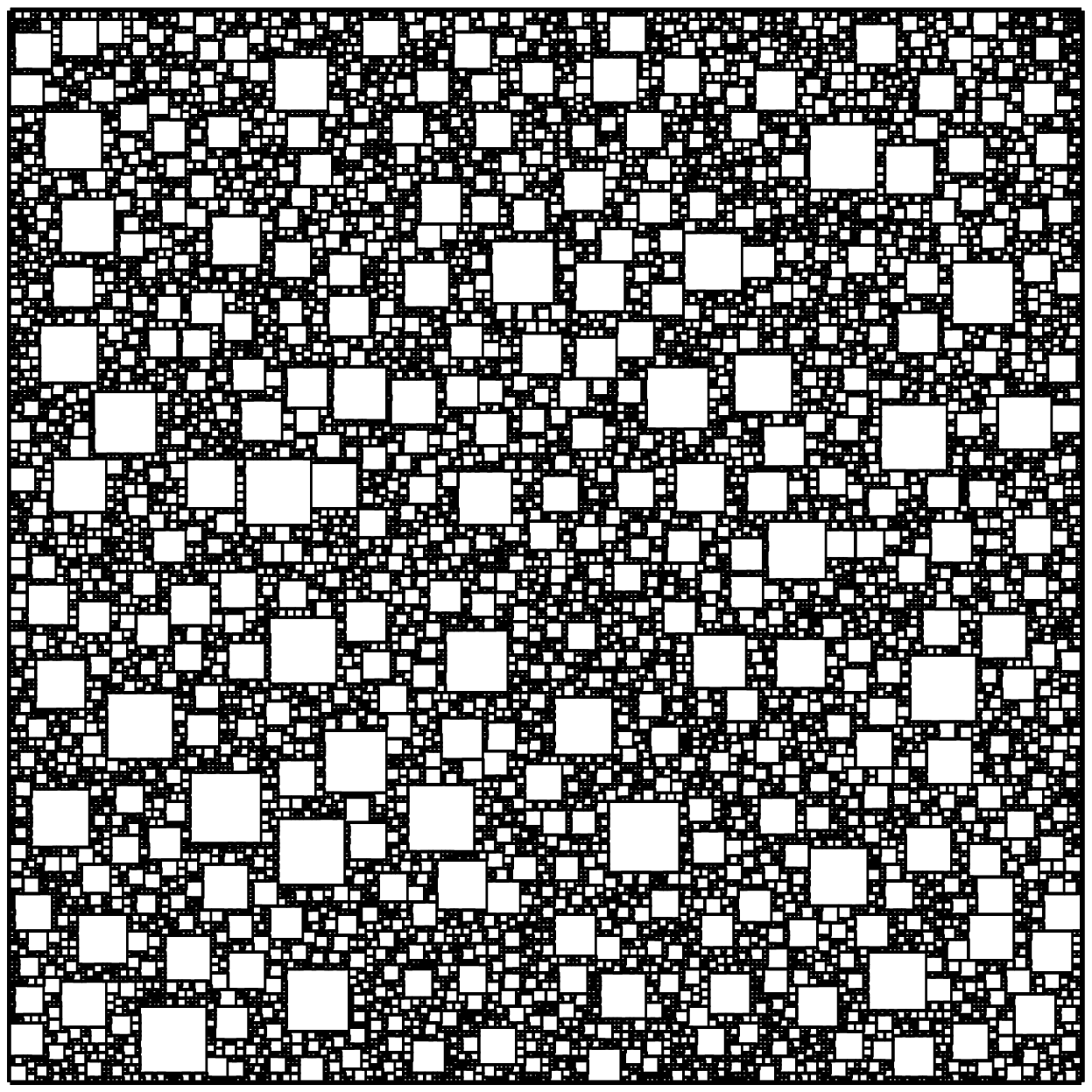}}\\
        3.6$\times$3.6 mm$^2$ \hspace{1in}
        5.3$\times$5.3 mm$^2$ \hfill
        9.0$\times$9.0 mm$^2$ \\
        700 Particles \hspace{1in} 2800 Particles \hfill
        11600 Particles \\
     \end{center}
 	\caption{Square particle model based on the 
		 size distribution of pressed PBX 9501.}
        \label{fig:pressedPBXsq}
  \end{figure}

  Of the various microstructures of PBX 9501 explored in the previous 
  section, the best estimate of the Young's modulus of PBX 9501 is provided 
  by the 100 particle model of the dry blend of PBX 9501 shown in 
  Figure~\ref{fig:dryBlendPBX}, overlaid by a 350$\times$350 grid.  The 
  effective properties of PBX 9501 shown in this section have been calculated 
  at different temperatures and strain rates using the 100 particle model of 
  Figure~\ref{fig:dryBlendPBX}.  Instead of the experimentally determined 
  values of moduli of HMX, data from molecular dynamics (MD) 
  simulations~\cite{Sewell99} have been used because estimates using MD
  data of the Young's modulus of PBX 9501 at room temperature and low strain 
  rate are slightly closer to the experimental value than estimates calculated 
  using the experimentally determined modulus of HMX.  

  The Young's modulus of HMX from MD simulations is 17,700 MPa and the 
  Poisson's ratio is 0.21 compared to a Young's modulus of 15,300 MPa and a 
  Poisson's ratio of 0.32 from experimental data.  The Young's modulus of 
  HMX is assumed constant at the temperatures and strain rates considered. 

  The PBX 9501 binder data collected from various 
  sources~\cite{Gray98,Dick98,Wetzel99,Cady00} are listed in 
  Table~\ref{tab:pbx9501input}.  
  The data are sorted in order of decreasing temperature followed by increasing
  strain rate.  As shown in Table~\ref{tab:pbx9501input}, the Young's modulus 
  of the binder generally increases with increase in strain rate and with 
  decrease in temperature though some of the available data do not follow this 
  trend.  Since values for the Poisson's ratio of the binder are not 
  available, a value of 0.49 has been used for all cases.  However, the value 
  is likely to be lower near the glass transition temperature of estane 
  (around -33$^o$C).  As before, a dirty binder has been used for the grid 
  cells not fully occupied by particles.  The differential effective medium 
  approximation~\cite{Markov00} has been used to calculate the effective 
  properties of the dirty binder using the moduli of HMX and the 
  binder.  Table~\ref{tab:pbx9501input} also shows the dirty binder properties 
  used in the simulations. It is interesting to note that the Poisson's ratio 
  of the dirty binder does not change significantly even though the Young's 
  modulus can vary considerably depending on the modulus of the PBX 9501 
  binder.

  Table~\ref{tab:pbx9501350Enu} shows the effective properties of the model 
  RVE computed using finite elements along with the available experimental 
  data on PBX 9501.  
  The experimental data are within 15\% of the finite element estimates for 
  temperatures close to 25$^o$C and at low strain rates.  Good agreement is 
  also observed for the data near 16$^o$C and at high strain rates.  However, 
  at 0$^o$C and a strain rate of 2200/s the experimental value is around 2 
  times the predicted values of Young's modulus while at -20$^o$C and high 
  strain rate, the experimental value is half the estimate.  Experimental data 
  on the Poisson's ratio of PBX 9501 are available only at 25$^o$C and 0.005/s 
  strain rate.  The predicted Poisson's ratio under these conditions is 
  considerably lower that the experimental value.  Since the Poisson's ratio 
  is highly sensitive to numerical error~\cite{Garbo02}, more accurate 
  numerical calculations may be required to resolve this issue.
  
  These results show that, based on existing data on PBX 9501, it is not 
  possible to conclude that the effective moduli of PBX 9501 can be predicted 
  accurately if the moduli of its constituents are known for various 
  temperatures and strain rates.  However, the results do show that for 
  certain RVEs, close approximations to the effective Young's modulus of 
  PBX 9501 can be obtained from two-dimensional finite element simulations.  
 
  \section{Summary and conclusions}
  A two-dimensional finite element based method for approximating the 
  effective elastic moduli of particulate composites has been presented.  This 
  approach has been applied to representative volume elements (RVEs) 
  containing circular particles for volume fractions from 0.1 to 0.92.  The 
  estimated Young's moduli and Poisson's ratios have been found to closely 
  approximate those from differential effective medium 
  approximations.  Comparisons with three-dimensional finite element 
  simulations have also shown that the two-dimensional approach provides good 
  approximations for RVEs containing polydisperse spherical particles.  The 
  estimates of Young's modulus from three-dimensional models have been found 
  to be slightly higher than from two-dimensional models.  On the other hand, 
  the Poisson's ratios from three-dimensional models have been found to be 
  lower than estimates from two-dimensional models.  

  Since polymer bonded explosives such as PBX 9501 contain two widely 
  separated particle sizes, a number of manually generated microstructures of 
  PBX 9501 have been simulated.  The models were designed to contain a few 
  large particles and a larger number of smaller interstitial particles while 
  avoiding particle-particle contact.  Such models have been found to 
  underestimate the Young's modulus of PBX 9501 considerably when discretized 
  with triangular elements.  However, when the same models are discretized 
  with square elements, considerably different estimates of Young's modulus 
  are obtained depending on the degree of discretization.  Hence, it is 
  extremely difficult to determine an appropriate RVE for high modulus 
  contrast and high volume fraction particulate composites that has the 
  optimal size, number of particles, particle shapes and size 
  distributions.  The RVE also has to be such that it can be easily 
  discretized using elements that are accurate and relatively computationally 
  inexpensive.

  Randomly generated microstructures based on the particle size distribution 
  of PBX 9501 show that the size of the RVE plays a lesser role in the 
  determination of the effective Young's modulus than the amount of 
  discretization used, even for highly refined meshes.  If aligned square 
  particles are used instead of circular particles, the predicted moduli 
  are considerably higher than the experimental moduli because contacts 
  between particles increase dramatically.  It is also observed that the 
  numerical simulations consistently underestimate the effective Poisson's 
  ratio.  Since the Poisson's ratio is highly sensitive to numerical 
  error, accurate numerical simulations have to be performed to resolve this 
  issue.
  
  Simulations of PBX 9501 for various temperatures and strain rates suggest 
  that if a RVE is chosen that predicts a reasonable value of Young's modulus 
  at a given temperature and strain rate, it can be used to obtain acceptable 
  estimates at other temperatures and strain rates.  Further experimental data 
  are required to confirm this finding.

  \section*{Acknlowledgements}
  This research was supported by the University of Utah Center for the
  Simulation of Accidental Fires and Explosions (C-SAFE), funded by the     
  Department of Energy, Lawrence Livermore National Laboratory, under 
  subcontract B341493.  

  \appendix
  \section{Two- and Three-Dimensional Moduli}\label{app:2D3D}
  The determination of the effective properties of random particulate 
  composites using numerical techniques requires the use of a three-dimensional
  representative volume element (RVE).  If a two-dimensional RVE is selected, 
  any plane strain representation immediately implies that the material is 
  composed of unidirectional cylindrical particles.  On the other hand, if a 
  plane stress representation is used, the physical interpretation of the 
  composite is a thin sheet cut from the particulate composite.

  One of the intents of this work has been to show that acceptable estimates
  of the effective elastic moduli of particulate composites can be obtained 
  from two-dimensional plane strain calculations of the stress and strain 
  fields.  However, two-dimensional calculations in a single plane of the 
  composite lead to two-dimensional or planar effective elastic moduli.  For 
  example, if a uniform uniaxial displacement $u_1$ is applied to a unit 
  square of an isotropic, homogeneous, linear elastic material with Young's 
  modulus $E$ and Poisson's ratio $\nu$ in plane strain, the resulting 
  reaction force is $F_1$ and the displacement perpendicular to the direction
  of the applied displacement is $-u_2$.  Since the square is of unit size, the
  nonzero strains are $\epsilon_{11} = u_1$ and $\epsilon_{22} = -u_2$.  The 
  nonzero stresses are $\sigma_{11} = F_1$ and $\sigma_{33} = \nu F_1$, the 
  latter obtained from the constitutive relation
  \begin{equation}
    \epsilon_{ij} = (1/E)[(1+\nu)\sigma_{ij} - \nu\sigma_{kk}\delta_{ij}]
     ~~i,j,k=1,2,3.
  \end{equation}
  Note that $E$ and $\nu$ are three-dimensional or true linear elastic moduli.
  It can be shown that the apparent Poisson's ratio in the plane 
  $(\epsilon_{22}/\epsilon_{11})$ is given by
  \begin{equation}
    \nu_{2D} = \epsilon_{22}/\epsilon_{11} = u_2/u_1 = \nu/(1-\nu) 
     \label{eq:2D3D_nu}
  \end{equation}
  and the apparent Young's modulus in the plane $(\sigma_{11}/\epsilon_{11})$ 
  is given by
  \begin{equation}
  E_{2D} = \sigma_{11}/\epsilon_{11} = F_1/u_1 = E/(1-\nu^2) \label{eq:2D3D_E}
  \end{equation}
  These plane moduli, $E_{2D}$ and $\nu_{2D}$, are referred to as 
  two-dimensional moduli following Jun and Jasiuk~\cite{JunJas93} and 
  Torquato~\cite{Torq01}.  The reason for such a name is that these apparent 
  moduli are the same as those obtained if the constitutive equations are 
  expressed, in two-dimensional form, as
  \begin{equation}
  \epsilon_{ij} = (1/E_{2D})[(1+\nu_{2D})\sigma_{ij} - \nu_{2D}\sigma_{kk}
     \delta_{ij}]~~i,j,k=1,2.
  \end{equation}

  The results presented in this work suggest that these relations between 
  two- and three-dimensional elastic moduli (which are exact only for 
  isotropic, homogeneous, linear elastic materials) can be used to obtain
  reasonable approximations of three-dimensional elastic moduli of random 
  particulate composites from two-dimensional numerical simulations.


  \newpage
  \section*{Figure captions}
  Figure 1.  RVEs containing 10\% to 92\% by volume of circular particles.
              $f_p$ is the volume fraction of particles in a RVE.\\
  Figure 2.  Comparison of finite element (FEM) and differential
              effective medium (DEM) predictions.  $E_b$ is the Young's 
              modulus of the binder.\\
  Figure 3.  Two- and three-dimensional RVEs containing 70\%, 75\% and 
              80\% particles by volume. $f_p$ is the volume fraction of 
              particles.\\
  Figure 4.  Two- and three-dimensional estimates of effective Young's modulus
	      and Poisson's ratio.\\
  Figure 5.  Microstructure of PBX 9501 (adapted from ~\cite{Skid98}).
              The largest particles are around 300 microns in size.\\
  Figure 6.  Manually generated microstructures containing $\sim$ 90\% 
              by volume of circular particles.\\
  Figure 7.  Microstructure containing 92\% particles modeled with 
              six-noded triangles and four-noded squares.\\
  Figure 8.  Circular particle models based on the dry blend of PBX 9501. \\
  Figure 9.  Circular particle models based on the 
	      particle size distribution of pressed PBX 9501. \\
  Figure 10. Square particle model based on the 
	      size distribution of pressed PBX 9501.

  \clearpage
  \section*{Tables}
  \clearpage
  \begin{table}
    \caption{Effective moduli of the six model PBX 9501 microstructures
	     from FEM calculations six-noded triangular elements.
             $E$ is the Young's modulus and $\nu$ is the Poisson's ratio.} 
    \label{tab:pbxManualFEM}
    \medskip
    \begin{center}
    \begin{tabular}{lccccccccc}
      \hline
      & Expt. &\multicolumn{6}{c}{Model RVE} & Mean & Std. Dev. \\
      &       & 1 & 2 & 3 & 4 & 5 & 6 &  & \\
      \hline
      E (MPa) & 1013 & 116 & 126 & 130 & 42 & 183 & 192 & 132 & 54 \\
      $\nu$  & 0.35 & 0.34 & 0.32 & 0.32 & 0.44 & 0.28 & 0.25 & 0.33 & 0.07 \\
      \hline
    \end{tabular}
    \end{center}
  \end{table}
  \clearpage
  \begin{table}
    \caption{Effective elastic moduli of the four models of the dry blend 
             of PBX 9501.}\medskip
    \label{tab:dryBlendPBX}
    \begin{center}
    \begin{tabular}{llcccccc}
    \hline
    No. of    & Size & \multicolumn{3}{c}{Young's Modulus (MPa)} &
			     \multicolumn{3}{c}{Poisson's Ratio} \\
         Particles & (mm) & \multicolumn{2}{c}{FEM} & Expt. & 
                             \multicolumn{2}{c}{FEM} & Expt. \\
                  &      & 256$\times$256 & 350$\times$350 &  & 
                             256$\times$256 & 350$\times$350 &  \\
    \hline
    100 & 0.65 & 1959 & 968  & 1013 & 0.22 & 0.20 & 0.35 \\
    200 & 0.94 & 2316 & 1488 & 1013 & 0.23 & 0.23 & 0.35 \\
    300 & 1.13 & 2899 & 2004 & 1013 & 0.25 & 0.24 & 0.35 \\
    400 & 1.33 & 4350 & 2845 & 1013 & 0.25 & 0.25 & 0.35 \\
    \hline
    \end{tabular}
    \end{center}
  \end{table}
  \clearpage
  \begin{table}
    \caption{Material properties of the dirty binder for the four pressed 
	     PBX 9501 microstructures. $f_p$ is the volume fraction of 
             particles.} \medskip
        \label{tab:pressedPBX}
         \begin{center}
	\begin{tabular}{lcccc}
	   \hline
	   No. of    & $f_p$ &  $f_p$    &  
		\multicolumn{2}{c}{Properties of dirty binder} \\
	   Particles &       & in binder &  E (MPa)  & $\nu$ \\
	   \hline
	   100 & 0.89 & 0.28 & 1.583 & 0.484 \\
	   200 & 0.87 & 0.37 & 2.1395 & 0.481 \\
	   500 & 0.86 & 0.43 & 2.713 & 0.478 \\
	   1000 & 0.855 & 0.45 & 2.952 & 0.477 \\
	   \hline
        \end{tabular}
         \end{center}
  \end{table}
  \clearpage
  \begin{table}
    \caption{Effective elastic moduli of the four models of pressed PBX 9501.}
    \medskip
    \label{tab:pressedPBXCircle}
    \begin{center}
    \begin{tabular}{llcccccc}
    \hline
    No. of    & Size & \multicolumn{3}{c}{Young's Modulus (MPa)} &
			     \multicolumn{3}{c}{Poisson's Ratio} \\
         Particles & (mm) & \multicolumn{2}{c}{FEM} & Expt. & 
                             \multicolumn{2}{c}{FEM} & Expt. \\
                   &      & 256$\times$256 & 350$\times$350 &  & 
                             256$\times$256 & 350$\times$350 &  \\
    \hline
    100  & 0.36 & 2925 & 1798 & 1013 & 0.23 & 0.14 & 0.35 \\
    200  & 0.42 & 3342 & 2408 & 1013 & 0.21 & 0.20 & 0.35 \\
    500  & 0.54 & 5256 & 3994 & 1013 & 0.25 & 0.23 & 0.35 \\
    1000 & 0.68 & 6171 & 4756 & 1013 & 0.25 & 0.25 & 0.35 \\
    \hline
    \end{tabular}
    \end{center}
  \end{table}
  \clearpage
  \begin{table}
    \caption{Material properties of the dirty binder for the three 
		 PBX microstructures with square particles.} \medskip
        \label{tab:pressedPBXsq}
        \begin{center}
	\begin{tabular}{lcccc}
	   \hline
	   No. of    & $f_p$ &  $f_p$    &  
		\multicolumn{2}{c}{Properties of dirty binder} \\
	   Particles &       & in binder &  E (MPa)  & $\nu$ \\
	   \hline
	   700 & 0.868 & 0.39 & 2.358 & 0.4799 \\
	   2800 & 0.866 & 0.40 & 2.448 & 0.4795 \\
	   11600 & 0.863 & 0.42 & 2.588 & 0.4788 \\
	   \hline
        \end{tabular}
        \end{center}
  \end{table}
  \clearpage
  \begin{table}
    \caption{Effective stiffnesses of the model microstructures 
                 with square particles.}\medskip
    \label{tab:pressedPBXSquare}
        \begin{center}
    \begin{tabular}{llcccc}
    \hline
    No. of    & Size & \multicolumn{2}{c}{Young's Modulus (MPa)} &
			     \multicolumn{2}{c}{Poisson's Ratio} \\
    Particles & (mm) & FEM (256$\times$256) & Expt. & 
                             FEM (256$\times$256) & Expt. \\
    \hline
    700   & 3.6 & 9119 & 1013 & 0.26 & 0.35 \\
    2800  & 5.3 & 9071 & 1013 & 0.27 & 0.35 \\
    11600 & 9.0 & 9593 & 1013 & 0.27 & 0.35 \\
    \hline
    \end{tabular}
        \end{center}
  \end{table}
  \clearpage
  \begin{table}
    \caption{Material properties used to calculate the effective moduli of 
             PBX 9501 for different temperatures and strain rates.  Numbers 
             in square brackets refer to the source of the binder data.} 
    \medskip
    \label{tab:pbx9501input}
    \begin{tabular}{cccccccc}
      \hline
      Temp.& Strain & \multicolumn{2}{c}{HMX} & 
                      \multicolumn{2}{c}{PBX 9501 Binder} &
		      \multicolumn{2}{c}{'Dirty' Binder} \\
      & Rate & Young's & Poisson's & Young's & Poisson's & Young's & 
                         Poisson's \\
      & & Modulus & Ratio & Modulus & Ratio & Modulus & Ratio \\
      ($^o$C)&(/s)&(MPa)   &  & (MPa) &  & (MPa) & \\
      \hline
      25 & 0.005 & 17700 & 0.21 & 0.59~\cite{Wetzel99} & 0.49 & 2.15 & 0.479 \\
      25 & 0.008 & 17700 & 0.21 & 0.73~\cite{Wetzel99} & 0.49 & 2.66 & 0.479 \\
      25 & 0.034 & 17700 & 0.21 & 0.81~\cite{Wetzel99} & 0.49 & 2.95 & 0.479 \\
      25 & 0.049 & 17700 & 0.21 & 0.82~\cite{Wetzel99} & 0.49 & 2.99 & 0.479 \\
      25 & 2400  & 17700 & 0.21 & 300~\cite{Dick98}    & 0.49 & 1001 & 0.474 \\
      22 & 0.001 & 17700 & 0.21 & 0.47~\cite{Cady00}   & 0.49 & 1.71 & 0.479 \\
      22 & 1     & 17700 & 0.21 & 1.4~\cite{Cady00}    & 0.49 & 5.10 & 0.479 \\
      22 & 2200  & 17700 & 0.21 & 3.3~\cite{Cady00}    & 0.49 & 12.0 & 0.479 \\
      16 & 1700  & 17700 & 0.21 & 22.5~\cite{Gray98}   & 0.49 & 81.5 & 0.479 \\
       0 & 0.001 & 17700 & 0.21 & 0.85~\cite{Cady00}   & 0.49 & 3.10 & 0.479 \\
       0 & 1700  & 17700 & 0.21 & 246~\cite{Gray98}    & 0.49 & 833  & 0.475 \\
       0 & 2200  & 17700 & 0.21 & 4~\cite{Cady00}      & 0.49 & 14.6 & 0.479 \\
     -15 & 0.001 & 17700 & 0.21 & 1.4~\cite{Cady00}    & 0.49 & 5.10 & 0.479 \\
     -15 & 1     & 17700 & 0.21 & 5.7~\cite{Cady00}    & 0.49 & 20.7 & 0.479 \\
     -15 & 1000  & 17700 & 0.21 & 1600~\cite{Cady00}   & 0.49 & 4082 & 0.458 \\
     -20 & 0.001 & 17700 & 0.21 & 1.6~\cite{Cady00}    & 0.49 & 5.83 & 0.479 \\
     -20 & 1200  & 17700 & 0.21 & 1600~\cite{Cady00}   & 0.49 & 4082 & 0.458 \\
     -20 & 1700  & 17700 & 0.21 & 1333~\cite{Gray98}   & 0.49 & 3556 & 0.461 \\
     -40 & 0.001 & 17700 & 0.21 & 5.7~\cite{Cady00}    & 0.49 & 20.7 & 0.479 \\
     -40 & 0.001 & 17700 & 0.21 & 5.3~\cite{Gray98}    & 0.49 & 19.3 & 0.479 \\
     -40 & 1300  & 17700 & 0.21 & 1000~\cite{Cady00}   & 0.49 & 2838 & 0.464 \\
      \hline
    \end{tabular}
  \end{table}
  \clearpage
  \begin{table}
    \caption{Effective Young's modulus from finite element simulations of a
             100 particle model of the dry blend of PBX 9501.
             Numbers in square brackets refer to
	     the source of the PBX 9501 data.} \medskip
    \label{tab:pbx9501350Enu}
    \begin{tabular}{cccccc}
      \hline
      Temp.& Strain & \multicolumn{2}{c}{FEM Estimates} & 
                      \multicolumn{2}{c}{Experiments} \\
           & Rate   & Young's & Poisson's & Young's & Poisson's \\
           &        & Modulus & Ratio     & Modulus & Ratio \\
      ($^o$C) &(/s) &(MPa) &  & (MPa) &  \\
      \hline
      25 & 0.005 & 1001  & 0.19 & 1040~\cite{Wiegand98}$^{\text a}$ & - \\
      25 & 0.01  & 1080  & 0.20 & 770~\cite{Dick98}  & - \\
         &       &       &      & 1020~\cite{Gray98}$^{\text b}$ & - \\
      25 & 0.034 & 1123  & 0.20 & -                  & - \\
      25 & 0.05  & 1129  & 0.20 & 1013~\cite{Wetzel99} & 0.35 \\
      25 & 2400  & 11273 & 0.31 & -                  & - \\
      22 & 0.001 & 926   & 0.17 & -                  & - \\
      22 & 1     & 1399  & 0.23 & -                  & - \\
      22 & 2200  & 2050  & 0.26 & -                  & - \\
      16 & 1700  & 4981  & 0.30 & 4650~\cite{Gray98}$^{\text c}$ & - \\
       0 & 0.001 & 1144  & 0.21 & -                  & - \\
       0 & 1700  & 10808 & 0.31 & -                  & - \\
       0 & 2200  & 2243  & 0.27 & 5480~\cite{Gray98} & - \\ 
     -15 & 0.001 & 1399  & 0.23 & -                  & - \\
     -15 & 1     & 2652  & 0.28 & -                  & - \\
     -15 & 1000  & 14602 & 0.28 & -                  & - \\
     -20 & 0.001 & 1481  & 0.23 & -                  & - \\
     -20 & 1200  & 14602 & 0.28 & -                  & - \\
     -20 & 1700  & 14291 & 0.29 & 6670~\cite{Gray98}$^{\text d}$ & - \\
     -40 & 0.001 & 2652  & 0.28 & -                  & - \\
     -40 & 0.001 & 2562  & 0.27 & -                  & - \\
     -40 & 1300  & 13778 & 0.29 & 12900~\cite{Gray98}$^{\text d}$ & - \\
      \hline
    \end{tabular} \\
    a - experimental data are for a strain rate of 0.001/s. \\
    b - experimental data are for a strain rate of 0.011/s and a 
        temperature of 27$^o$C. \\
    c - experimental data are for a strain rate of 2250/s and a 
        temperature of 17$^o$C. \\
    d - experimental data are for a strain rate of 2250/s. 
  \end{table}
  \clearpage

  \section*{Figures}
  \clearpage
  \clearpage
  \clearpage
  \clearpage
  \clearpage
  \clearpage
  \clearpage
  \clearpage
  \clearpage
  \clearpage
\end{document}